\documentclass[10pt,letterpaper]{article}

\usepackage{changepage}
\usepackage[utf8]{inputenc}
\usepackage{textcomp,marvosym}
\usepackage{fixltx2e}
\usepackage{amsmath,amssymb}
\usepackage{cite}
\usepackage{nameref,hyperref}
\usepackage[right]{lineno}
\usepackage{microtype}
\DisableLigatures[f]{encoding = *, family = * }
\usepackage{rotating}
\usepackage{color}
\usepackage{graphicx}
\usepackage{graphics}
\usepackage{soul}
\usepackage[normalem]{ulem}
\usepackage{booktabs}
\usepackage{subcaption}

\usepackage{tikz}
\usetikzlibrary{positioning}

\usepackage[aboveskip=1pt,labelfont=bf,labelsep=period,justification=raggedright,singlelinecheck=off]{caption}

\makeatletter
\renewcommand{\@biblabel}[1]{\quad#1.}
\makeatother

\date{}

\usepackage{color}

\begin{document}
\vspace*{0.35in}

% Title must be 150 characters or less
\begin{flushleft}
{\Large
\textbf\newline{A Longitudinal Analysis of the Public Perception of the Opportunities and Challenges of the Internet of Things}
}
\newline
% Insert Author names, affiliations and corresponding author email.
\\
Arkaitz Zubiaga\textsuperscript{1}*,
Rob Procter\textsuperscript{1,2},
Carsten Maple\textsuperscript{1,2}
\\
\bf{1} University of Warwick, Coventry, UK
\\
\bf{2} Alan Turing Institute, London, UK
%\\
%\textcurrency University of Warwick, Coventry, United Kingdom

* a.zubiaga@warwick.ac.uk

% Insert additional author notes using the symbols described below. Insert symbol callouts after author names as necessary.
% 
% Remove or comment out the author notes below if they aren't used.
%
% Primary Equal Contribution Note
% \Yinyang These authors contributed equally to this work.

% Additional Equal Contribution Note
% \ddag These authors also contributed equally to this work.

% \textcurrency b Insert current address of second author with an address update
% \textcurrency c Insert current address of third author with an address update

% \dag Deceased

% Group/Consortium Author Note
% \textpilcrow Insert Collaborative Author line here

\end{flushleft}
\section*{Abstract}
The Internet of Things (or IoT), which enables the networked interconnection of everyday objects, is becoming increasingly popular in many aspects of our lives ranging from entertainment to health care. While the IoT brings a set of invaluable advantages and opportunities with it, there is also evidence of numerous challenges that are yet to be resolved. This is certainly the case with regard to ensuring the cyber security of the IoT, and there are various examples of devices being hacked. Despite this evidence, little is known about the public perceptions of the opportunities and challenges presented by the IoT. To advance research in this direction, we mined the social media platform Twitter to learn about public opinion about the IoT. Analysing a longitudinal dataset of more than 6.7 million tweets, we reveal insights into public perceptions of the IoT, identifying big data analytics as the most positive aspect, whereas security issues are the main public concern on the negative side. Our study serves to highlight the importance of keeping IoT devices secure, and remind manufacturers that it is a concern that remains unresolved, at least insofar as the public believes.

% \linenumbers

\section*{Introduction}

The Internet of Things (IoT) is a concept that refers to the networked interconnection of everyday objects, which are often equipped with ubiquitous intelligence \cite{xia2012internet}. The use of IoT devices is becoming commonplace in our daily lives given the growing presence of WiFi and 4G-LTE Internet connectivity \cite{gubbi2013internet,reiter2014wireless}. The IoT presents numerous applications in different contexts, including \cite{gubbi2013internet}: (1) home, e.g. entertainment, health monitoring, (2) transport, e.g. traffic control, parking management, (3) community, e.g. environment monitoring, surveillance, and (4) national, e.g. defense, remote monitoring. The utility of the IoT can range from mere personal use at home to use in the industry as well as by doctors or carers for remote assistance \cite{osseiran2017internet}.

As well as being a valuable technology for remote and networked control of devices and data sources, the IoT also comes with the caveats of other Internet-connected devices: potential security and privacy issues linked to the use of those devices \cite{xu2014security,farooq2015critical,hwang2015iot,zhao2013survey}. The fact that data associated with IoT devices is sent through the Internet and stored in the cloud can make it vulnerable \cite{tellez2016improving} and can expose IoT devices to hackers \cite{stanislav2015hacking,miessler2015iot}. Security and privacy of IoT devices are especially crucial when they are being used with personal data associated with sensitive aspects such as health care \cite{islam2015internet}.

There is indeed a growing concern about the security and privacy issues brought about by the IoT \cite{jing2014security,granjal2015security,sadeghi2015security,Maple2017security}. Scientists are among those calling for new regulatory approaches that will enable attacks to be intercepted, data authenticated, access controlled and the privacy of customers guaranteed \cite{weber2010internet}. Further, \cite{atzori2010internet} highlight that while \textit{``the main strength of the IoT idea is the high impact it will have on several aspects of everyday-life and behavior of potential users''}, they also warn about the numerous issues that still need to be addressed: \textit{``many challenging issues still need to be addressed and both technological as well as social knots have to be untied before the IoT idea being widely accepted. Central issues are making a full interoperability of interconnected devices possible, providing them with an always higher degree of smartness by enabling their adaptation and autonomous behavior, while guaranteeing trust, privacy, and security.''} In this work we set out to explore how these three concepts, namely trust, privacy and security, are perceived and potentially questioned by the general public.

Attitudes and perceptions of IoT security were analysed by \cite{asplund2016attitudes} through surveying a group of actors in the energy, water and health care sectors. 18 representatives from 11 different organisations were interviewed as part of the study. Respondents in this study reported concerns about the lack of IT maturity and potential drawbacks of the IoT technology. The researchers, however, noticed a certain degree of lack of awareness among respondents, highlighting that two of the respondents had not even heard the term IoT before.

A number of studies and surveys have gathered the knowledge and opinions of experts, as well as the perceptions of actors of relevant sectors. However, the opinion of the general public has been studied to a much lesser extent. To the best of our knowledge, two studies have used social media to assess the public perceptions of the IoT. One of these two works is by \cite{joseph2017review}, who analysed a small dataset of 40K tweets covering two months in 2016, examining among others the most frequent users and hashtags in the dataset. Their analysis did not identify content associated with privacy and/or security, potentially owing to the limited size of the dataset. The other work is by \cite{bian2016mining}, who used larger datasets for their analysis, however, with gaps in the collection. They used a lexicon-based sentiment classification system and an LDA topic modelling approach to analyse different topics and sentiments associated with IoT. They did find that one of the topics identified in their LDA was indeed security, showing that the public has some concern about it; their study did not however further dig into the sentiments associated with these particular topics.

In this work, we perform a longitudinal analysis of public perceptions of the opportunities and challenges presented by the IoT; we use the social media platform Twitter as a data source to analyse posts from 2009 to 2016. We use a state-of-the-art sentiment analysis classifier to analyse the polarity of posts over time, and use a range of algorithms to enhance the dataset with additional dimensions to explore, including the country of origin of tweets and the gender of users, among others. Moreover, we also use an LDA topic modelling algorithm to identify the topics discussed in tweets associated with the IoT, which enables us to perform a fine-grained analysis of polarity by topic. By further examining these topics, we find that ``Big data \& tech'' is the largest emerging topic associated with the opportunities afforded by the IoT, whereas the key negative aspect and the second by number of tweets is ``Security'', which we observe has been a constant concern since 2009. Our study reveals insight into public perception of the IoT, highlighting the key opportunities and challenges posed by this technology. The use of unsupervised approaches to conduct our study also enables the application of our analysis to analyse the public perception associated with other major topics discussed on Twitter.

\section*{Materials and Methods}

The present study was approved by the Warwick University Humanities \& Social Sciences Research Ethics Committee (HSSREC), which received full ethics approval with a duration of 48 months (ref 69/13-14, date: 30.05.2014), including approval to store, analyse and publish extracts from social media datasets.

In this study, we use the microblogging platform Twitter as the data source to mine public comments associated with the IoT. Other platforms were also considered at the beginning of this study, including, for example, Reddit.  These other potential sources were deemed unsuitable as the vast majority of posts were from \emph{tech savvy} people who were dealing with technical challenges of setting up and using IoT devices, rather than more general comments from users who would be more representative of the 'average' consumer. Rather, Twitter was found to be much more suitable, as there is a variety of users of different levels of technical sophistication, discussing a wide range of issues.

Using a combination of Twitter's search interface and its REST API, we harvested a collection of tweets spanning eight years between 1st January, 2009 and 31st December, 2016. Given that Twitter's API does not give access to old tweets, we scraped the tweets from the site's search engine, which allows going back to the oldest tweets. This scraping process enabled us to collect all tweet IDs that matched our search query; these tweet IDs were then used to retrieve all tweets and metadata from the REST API. Tweets that included one of the following keywords were collected: `\#iot', `internet of things', `\#internetofthings'. We filtered out non-English tweets, as well as retweets, leading to a dataset with 6,705,948 tweets. This large-scale dataset enabled longitudinal analysis of public comments associated with the IoT on Twitter.

To enable detailed analysis of this dataset by looking at additional factors that Twitter does not directly provide through its API, we pre-processed the dataset. This pre-processing included the following text mining, classification and further data collection steps that we carried out to complete the dataset:

\begin{itemize}
 \item \textbf{Unpacking of URLs:} Many of the links in tweets tend to be shortened by using URL shortening services such as bit.ly or tinyurl.com, which is a useful service for users, owing to the limited number of characters allowed in each tweet. Since we are interested in analysing the links that people direct to when discussing IoT, we unpacked the shortened URLs by using the cURL library.\footnote{\url{https://curl.haxx.se/}} In this process of unpacking URLs, we stored the ultimate URL to which a shortened URL directs to. While we also retrieved the content of the URLs, this is not used for our quantitative analysis, which focuses on the analysis of final URLs.
 
 \item \textbf{Classification of tweets by country of origin:} We used a state-of-the-art tweet geo-location tool \cite{zubiaga2017towards} to classify tweets by country; it is a multinomial logistic regression that combines a range of tweet content and metadata for the classification, which is publicly available.\footnote{\url{https://github.com/azubiaga/tweet-country-classification}} We used the publicly available, geolocated dataset with over 5 million tweets\footnote{\url{https://figshare.com/articles/Tweet_geolocation_5m/3168529}} for training the classifier, which then applied to our dataset of IoT tweets. This enabled us to analyse what people's opinions are with respect to the IoT in different countries. Given that only a small subset of the tweets analysed were geo-tagged, we evaluated the accuracy of the classifier over those tweets, i.e. tweets that are geolocated by the user's device. The results gave an overall 81.4\% accuracy in a classification task involving 217 countries.
 
 \item \textbf{Inferring gender of users:} To infer the gender of users we used the first name of the Twitter users utilising the SexMachine Python package\footnote{\url{https://pypi.python.org/pypi/SexMachine/}}. We extracted the first name of the author of a tweet from the ``name'' field of a tweet. With this first name as input, the Python package then returns one of male, female, mostly\_male, mostly\_female or andy (for unknown names). We only kept the gender label for those classified as male or female (i.e. excluding those classified as mostly\_male, mostly\_female), labelling the rest as unknown.
 
 \item \textbf{Sentiment analysis:} Each tweet was classified as positive, negative or neutral, using a state-of-the-art sentiment classifier \cite{wang2017tdparse} that determines the sentiment of a tweet with respect to IoT as a target. This is different from tweet-level sentiment classification, which classifies the overall sentiment expressed in the tweet, irrespective of the target. For instance, the tweet \textit{``I like that there is quite a lot of research on IoT devices lately''} bears a positive overall sentiment, however, the sentiment is neutral with respect to IoT. Target-specific sentiment classification is being increasingly used where the objective is to determine sentiment towards specific topics or entities \cite{wang2017totemss}.
\end{itemize}

Finally, we also processed the entire dataset for topic modelling using LDA \cite{blei2003latent}, for which we used the implementation provided with Gensim\footnote{\url{https://radimrehurek.com/gensim/}}. With LDA, we generated a set of 6 topics\footnote{Empirically, different numbers of topics ranging from 5 to 10 were tested to find the optimal number that would lead to as many distinct topics as possible while avoiding repeated topics. Six was found to be the optimal number of topics.} from our dataset, and we categorised each tweet in the dataset into one of these topics. Following the topic modelling based clustering approach proposed by Wang et al. \cite{wang2017hierarchical}, we summed the topic probabilities for each keyword in a tweet, choosing the topic that maximised this sum as the category for the tweet.

To validate that the assumption that the growth of Twitter from 2009 to 2016 Twitter data has not had a significant effect on the dataset, we also collected data from Google Trends. Using the search engine of Google Trends, we retrieved the trends of the ``Internet of Things'' as a topic between 2009 and 2016. With this, we built monthly activity counts for Google Trends, and we did the same for Twitter data. A comparison between these two lists of frequencies showed a high Pearson correlation value of $\rho = 0.8978$ ($p < 2.2e^{-16}$), suggesting the growth of mentions of IoT was commensurate with the growth on Google Trends, without a significant effect of Twitter's user base growth.

\section*{Results}

\subsection*{Sentiment analysis}

First, we examined the sentiment expressed by users from 2009 to 2016. Out of the 6,705,948 tweets in the dataset, 295,674 (4.4\%) have either positive or negative sentiment, whereas the rest have a neutral sentiment. This suggests that a vast majority of the tweets are likely to be neutral stories and other comments that do not express any sentiment towards IoT. To validate the output of our sentiment classifier and to make sure that it does not have a tendency to label tweets as neutral, we also tested an alternative sentiment classifier, Vader \cite{hutto2014vader}, a widely used sentiment classifier for tweets and social media. A comparison of labels predicted by our classifier and by Vader led to an overlap of 95.45\% of the instances, showing the prevalence of neutral tweets in the dataset. A closer look at a subset of the neutral tweets shows that in fact many of those tweets are linking to news articles, whose headlines are copied into the tweets and have neutral sentiment.

\begin{figure}[tbh]
  \begin{center}
   \includegraphics[width=1.0\textwidth]{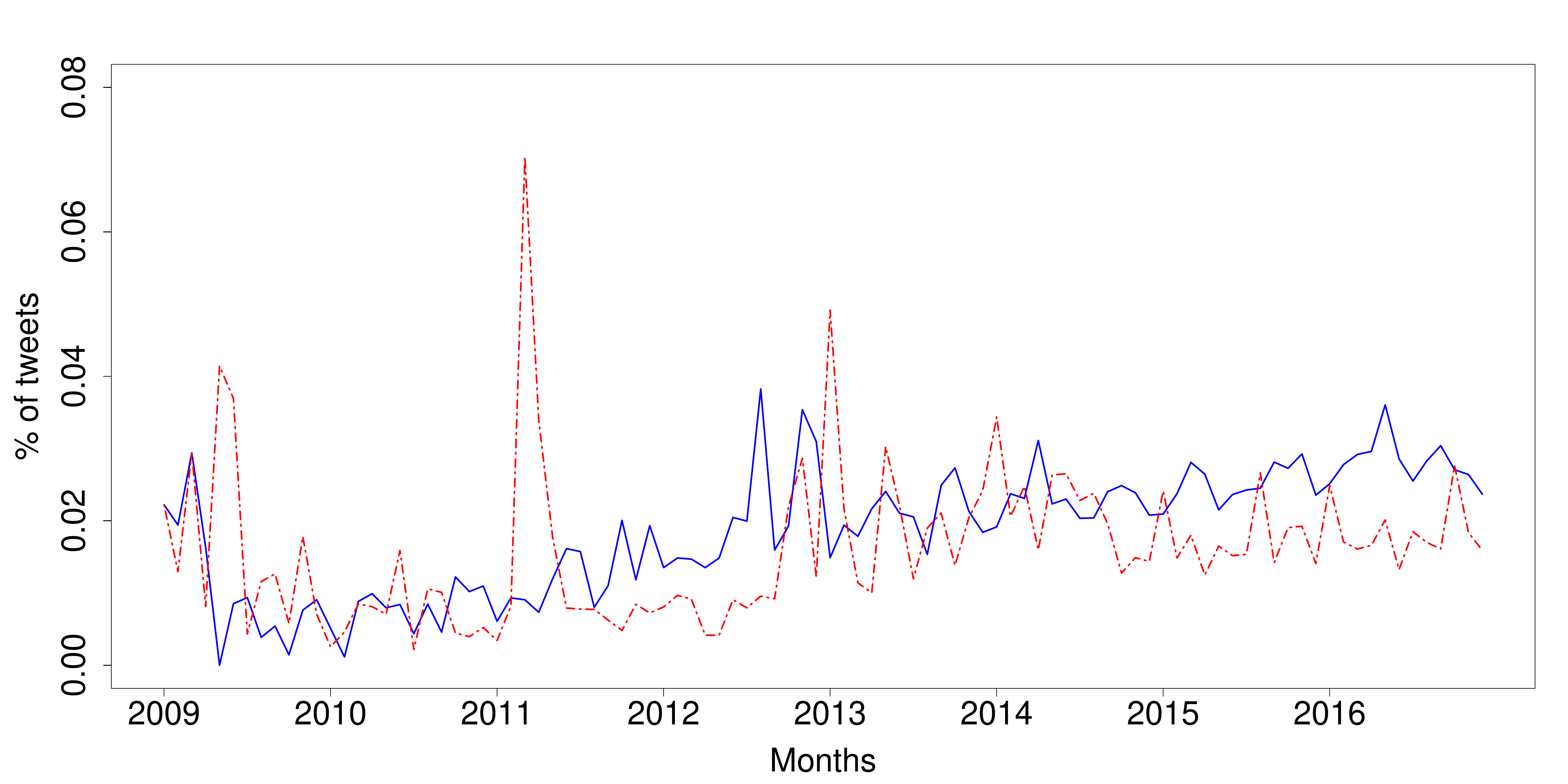}
   \caption{Sentiment of tweets discussing the Internet of Things over time (Jan 2009 - Dec 2016). Blue, solid line: positive, red, dashed line: negative.}
   \label{fig:sentiment}
  \end{center}
\end{figure}

Given that our objective was to look at the benefits and concerns of the IoT that people share opinions about, we focused our sentiment analysis on the positive and negative tweets. Figure \ref{fig:sentiment} shows a timeline with monthly percentages of positive and negative tweets over time. We observe that for the first years between 2009 and 2013, both positive and negative tweets went hand in hand, with very similar percentages. There are exceptions where negative tweets experienced significant spikes. This is potentially due to specific news stories that revealed negative aspects of the IoT, which we will analyse below. It is only in the last two years, 2015 and 2016, that the positive tweets exceeded the negative tweets quite consistently, suggesting that the positivity of the IoT is improving over time.

\begin{figure}[tbh]
  \begin{center}
   \includegraphics[width=1.0\textwidth]{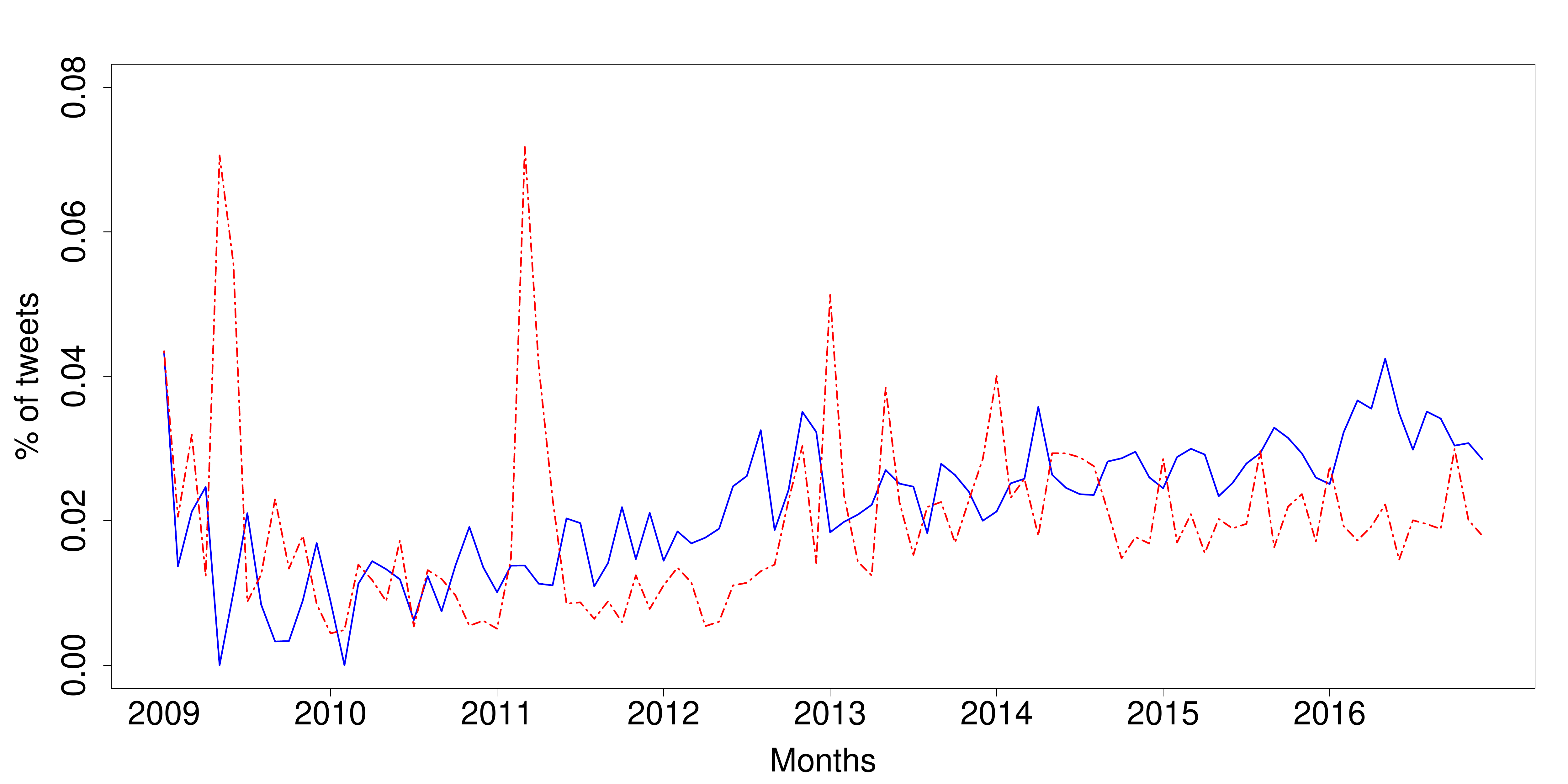}
   \caption{Sentiment of tweets posted by humans discussing the Internet of Things over time (Jan 2009 - Dec 2016). Blue, solid line: positive, red, dashed line: negative.}
   \label{fig:sentiment-humans}
  \end{center}
\end{figure}

One issue that arises from this sentiment analysis is that one may think that the results can be rigged by the presence of bots, i.e. automated accounts or 'bots' that tweet in volume positive or negative comments on the IoT \cite{ferrara2016rise}. This may happen, for instance, if a company wants to promote their IoT products (i.e. positive tweets) or competitors want to widely disseminate the drawbacks of IoT systems (i.e. negative tweets). To quantify and to validate our analysis, we examined the sentiment expressed by human users. To do this, we considered the subset of tweets for which the author had been identified as either male or female, i.e. having a human name. This led to a subsample with 2,602,138 tweets (38.8\% of the whole) that were identified as tweets by people. Figure \ref{fig:sentiment-humans} shows the sentiment expressed over time by human users. The overall sentiment trends in Figure \ref{fig:sentiment} and the human-only sentiment in Figure \ref{fig:sentiment-humans} have a Pearson correlation of 0.93 when we look at the positive tweets, and 0.95 for negative tweets, which can also be observed in the significant similarity between the charts. One striking difference that stands out is the sharper spikes seen in the negative tweets, which is more pronounced than in the positive tweets.

\subsection*{Topics}

\begin{figure}[p]
  \centering
  \begin{subfigure}[t]{0.48\textwidth}
   \centering
   \includegraphics[height=2in]{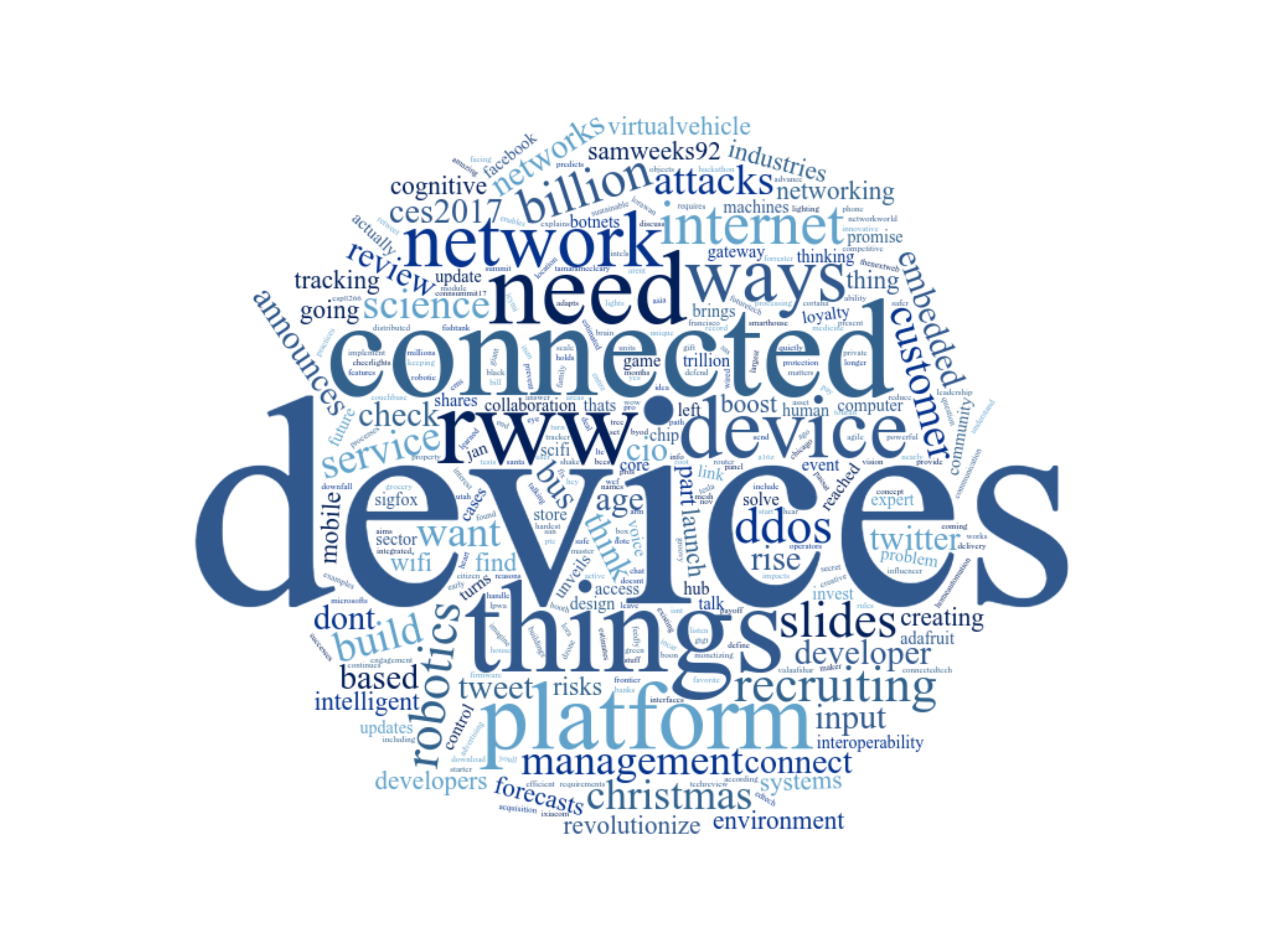}
   \caption{Devices (15.27\%)}
  \end{subfigure}
  ~
  \begin{subfigure}[t]{0.48\textwidth}
   \centering
   \includegraphics[height=2in]{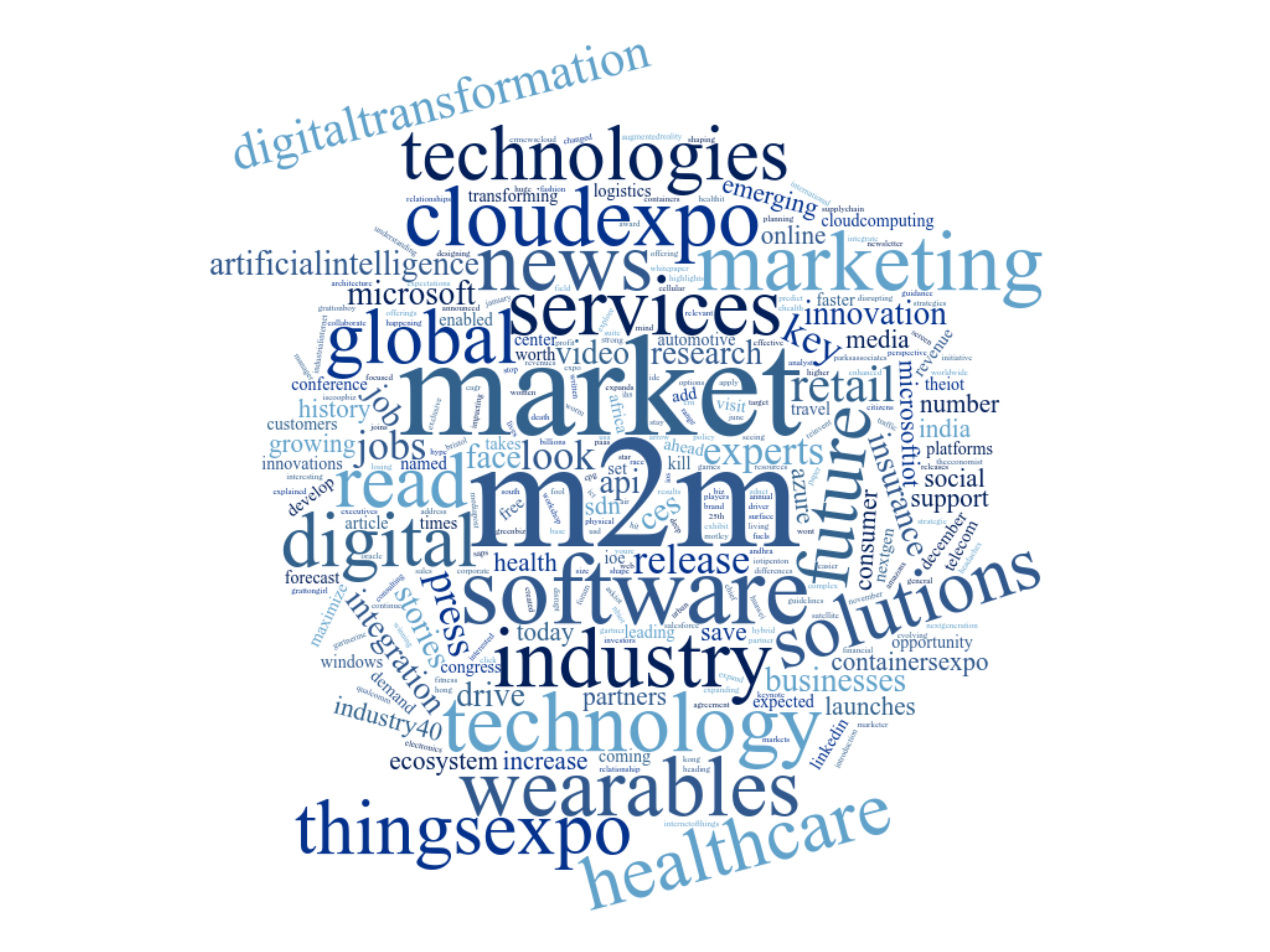}
   \caption{Machine to machine communication (16.14\%)}
  \end{subfigure}
  \\
  \begin{subfigure}[t]{0.48\textwidth}
   \centering
   \includegraphics[height=2in]{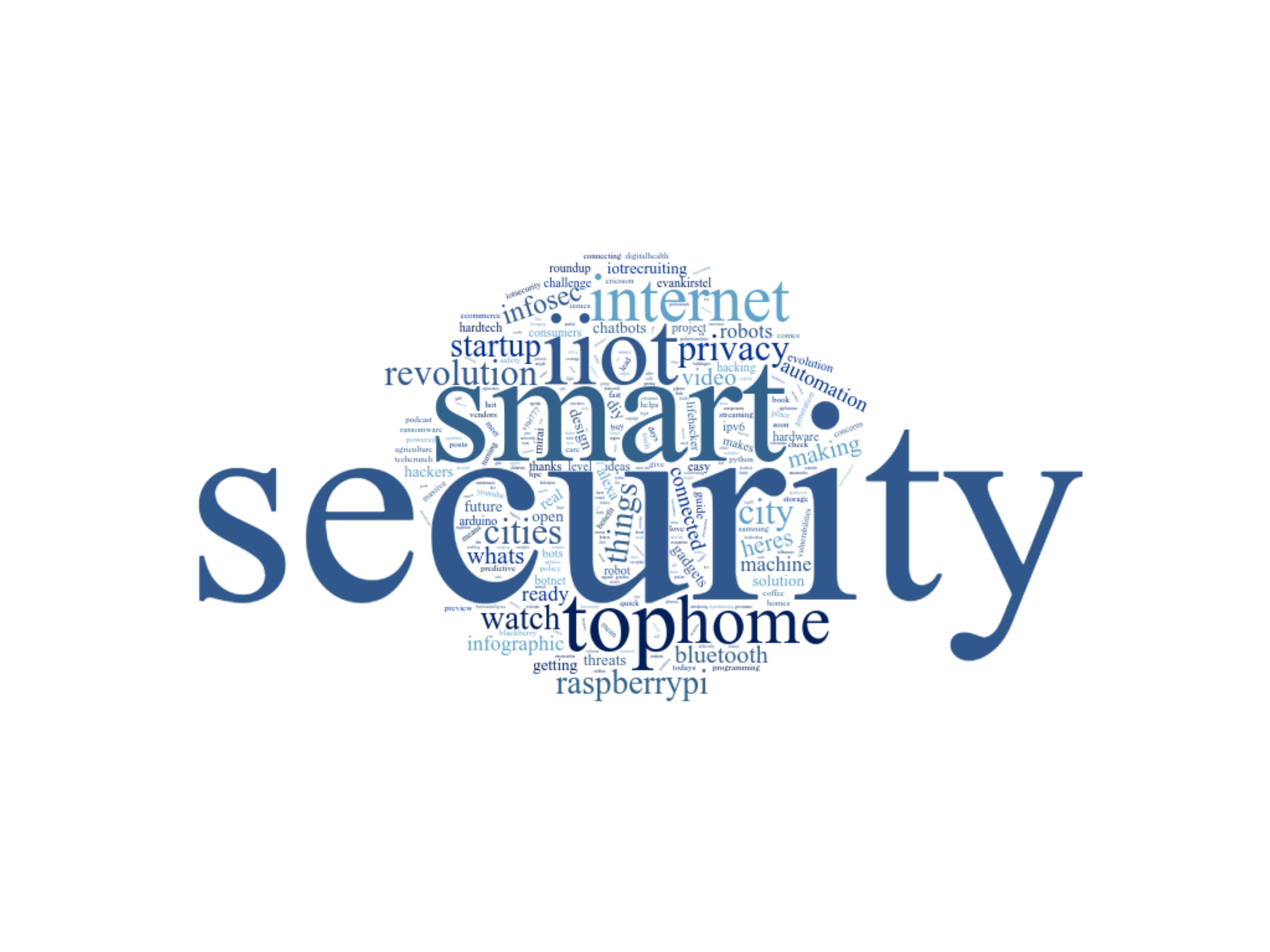}
   \caption{Security (20.49\%)}
  \end{subfigure}
  ~
  \begin{subfigure}[t]{0.48\textwidth}
   \centering
   \includegraphics[height=2in]{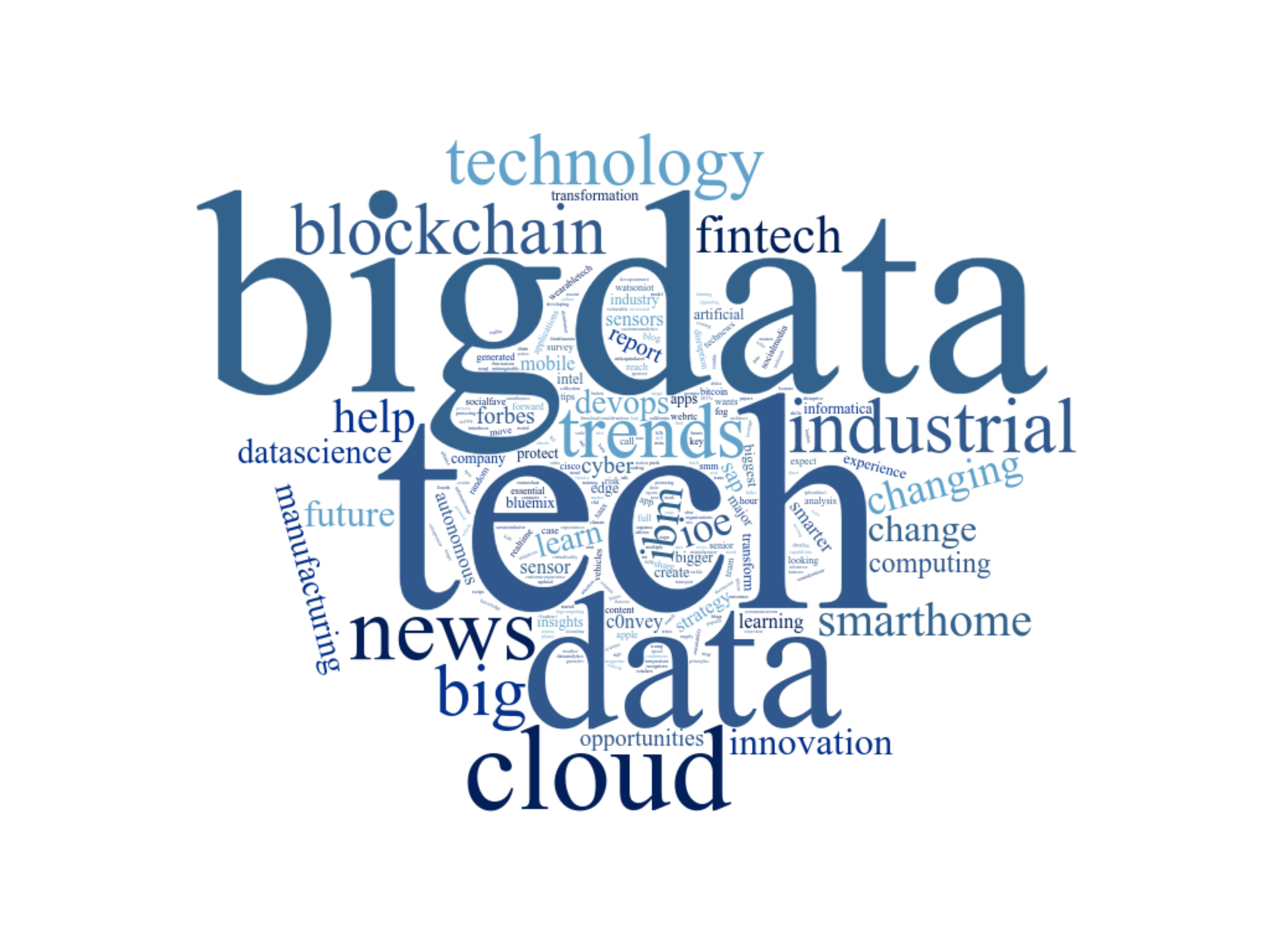}
   \caption{Big data \& Tech (29.92\%)}
  \end{subfigure}
  \\
  \begin{subfigure}[t]{0.48\textwidth}
   \centering
   \includegraphics[height=2in]{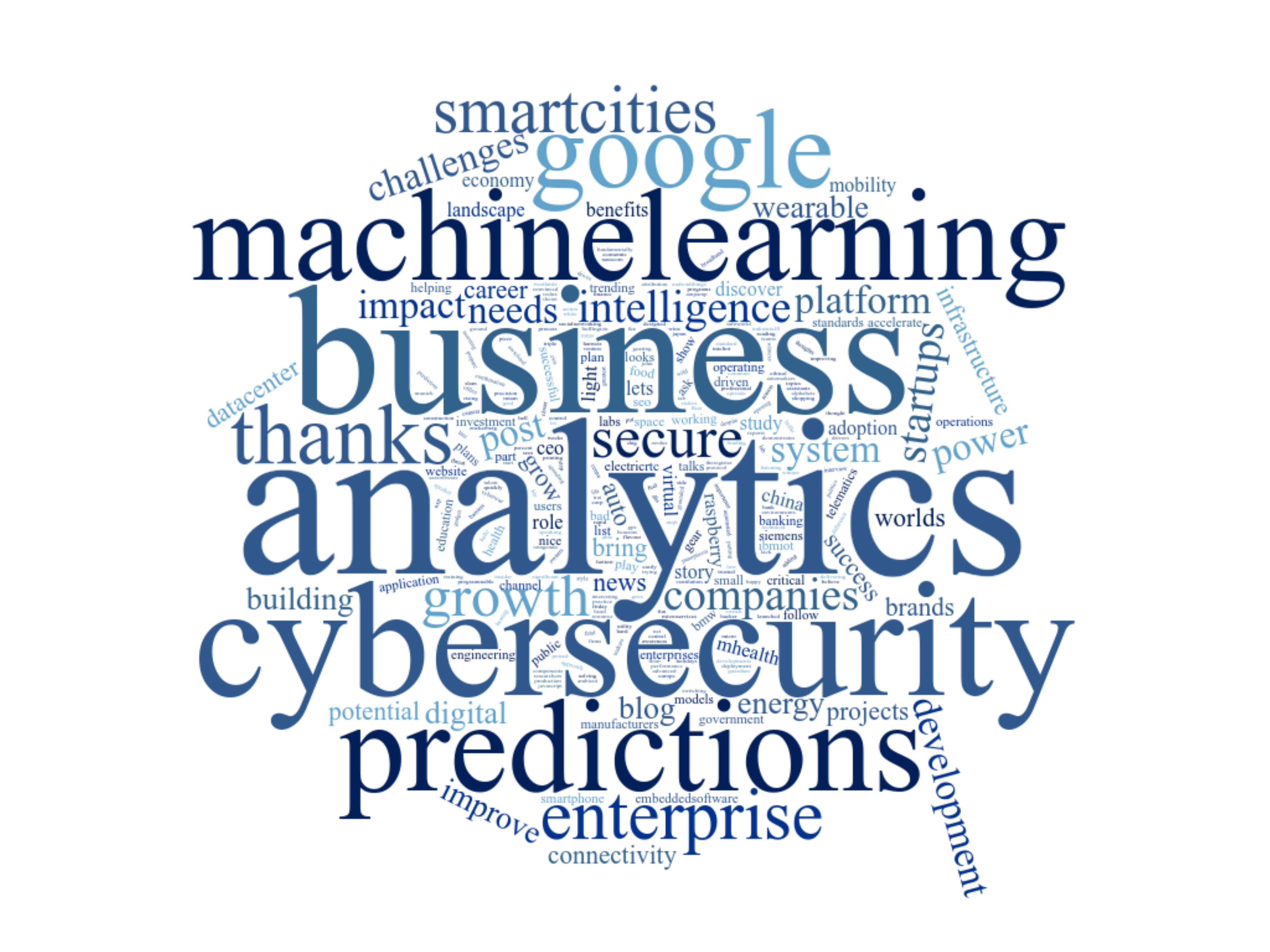}
   \caption{Analytics (12.09\%)}
  \end{subfigure}
  ~
  \begin{subfigure}[t]{0.48\textwidth}
   \centering
   \includegraphics[height=2in]{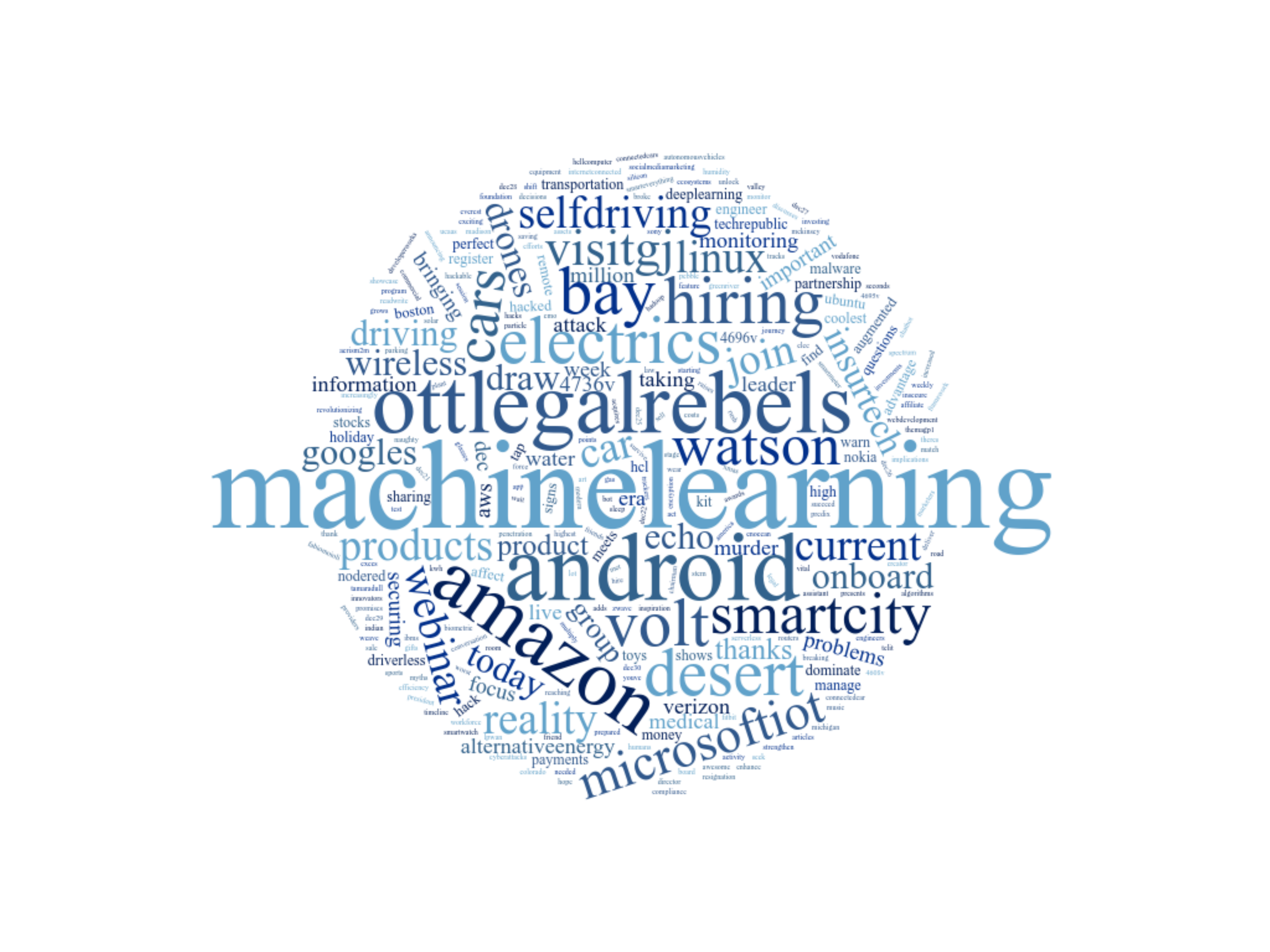}
   \caption{Machine learning (6.09\%)}
  \end{subfigure}
  \caption{Word clouds for topics extracted using LDA, as well as percentage of the whole represented by each topic.}
  \label{fig:lda}
\end{figure}

While the sentiment analysis above can be indicative of the overall perceptions of the IoT, we were interested in examining in detail the different issues and topics that people bring up around the IoT, and the sentiment expressed towards those topics. For this, we identified 6 different topics using a Latent Dirichlet Allocation algorithm (LDA), as described above. Figure \ref{fig:lda} shows the keywords representing the six topics, where the importance of each keyword is represented by its font size, with larger font sizes representing more important keywords. We observe that five of these six topics are associated with either positive aspects or business opportunities of the IoT, including analytics, machine learning, big data and tech, machine to machine communication, and devices. There is, however, a sixth topic, where one of the possible issues of the IoT is discussed, security. The largest topics in terms of the number of tweets associated with them are first,``Big data \& Tech'' with nearly 30\% of the tweets, and second,``Security'' with over 20\% of the tweets. This demonstrates that security is indeed a key concern around the IoT.

\begin{figure}[tbh]
  \begin{center}
   \includegraphics[width=1.0\textwidth]{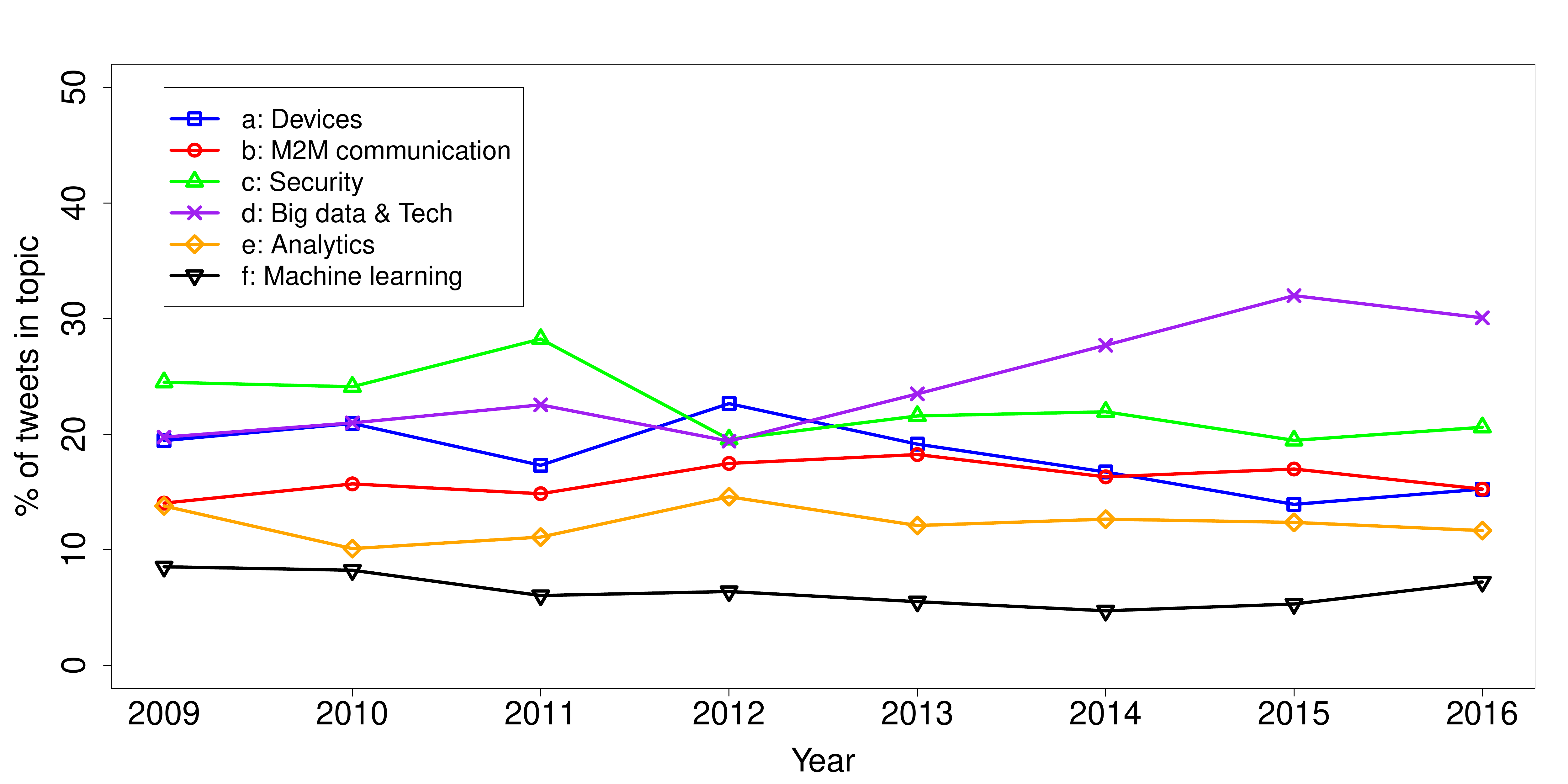}
   \caption{Popularity of each of the six topics over time.}
   \label{fig:lda-year}
  \end{center}
\end{figure}

To further drill down into the analysis and quantification of topics, a temporal trend of topic sizes is shown in Figure \ref{fig:lda-year}. The trends show that while ``Big data \& Tech'' is increasing in popularity.  The opportunities that the IoT is bringing to the big data and technology businesses is indeed becoming increasingly popular, as is reflected in this analysis. ``Security'' is stable over the last few years, with a slight drop since the 2009 and remains a big issue, consistently ranking second from 2013 to 2016.

\begin{figure}[htb]
  \centering
  \begin{subfigure}[t]{0.48\textwidth}
   \centering
   \includegraphics[height=1.2in]{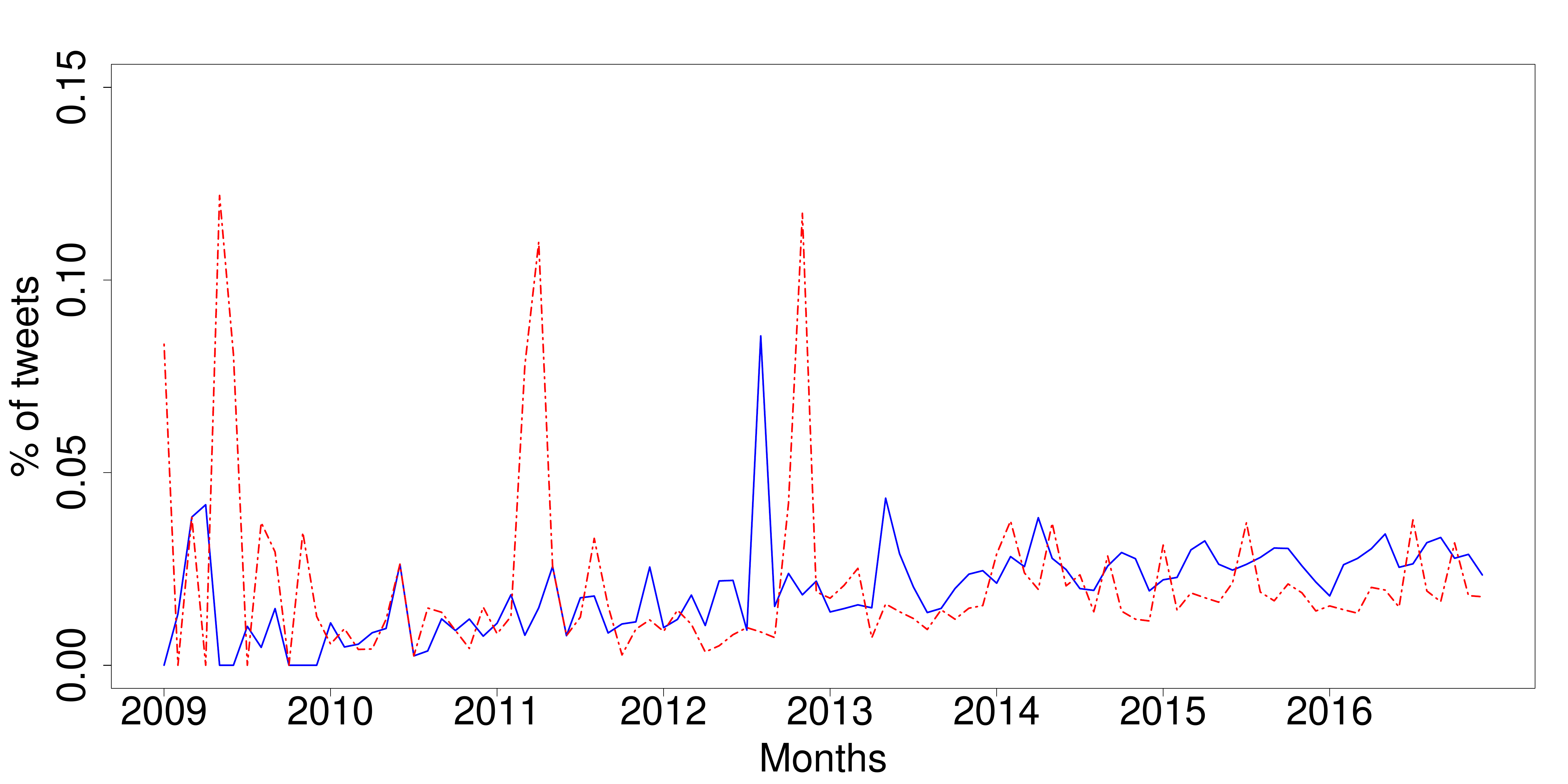}
   \caption{Sentiment for topic (a): Devices}
  \end{subfigure}
  ~
  \begin{subfigure}[t]{0.48\textwidth}
   \centering
   \includegraphics[height=1.2in]{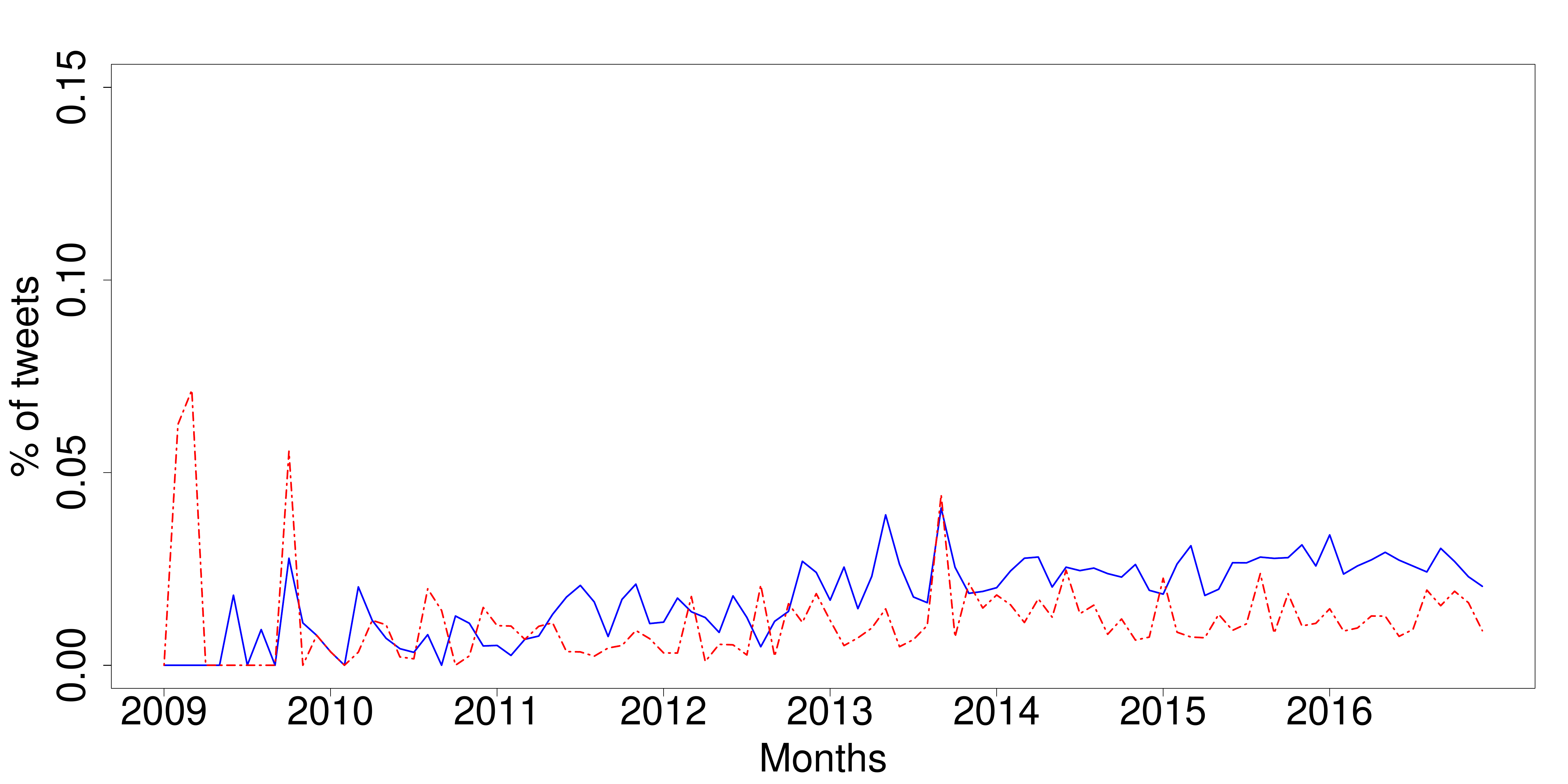}
   \caption{Sentiment for topic (b): M2M communication}
  \end{subfigure}
  \\
  \begin{subfigure}[t]{0.48\textwidth}
   \centering
   \includegraphics[height=1.2in]{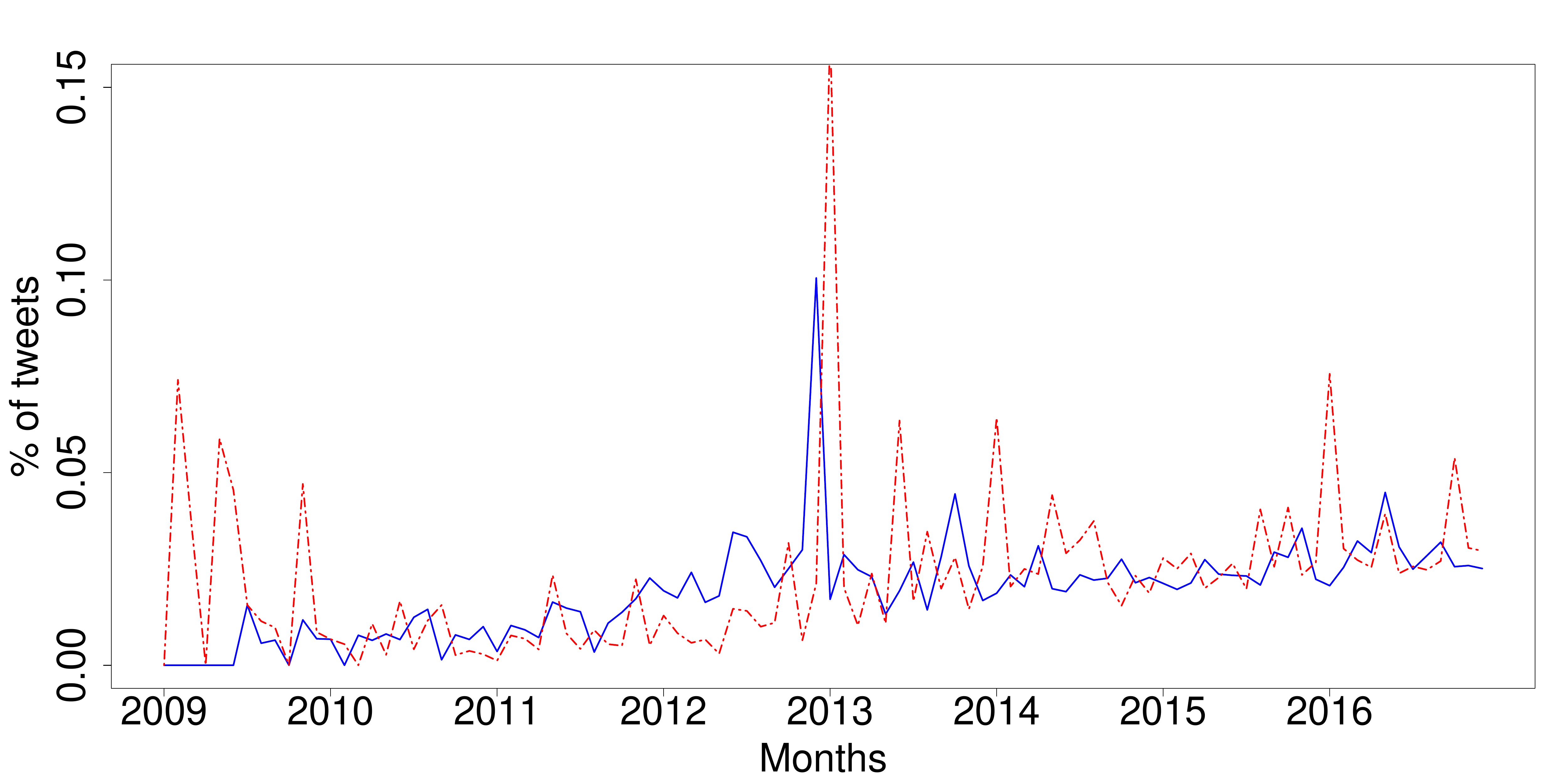}
   \caption{Sentiment for topic (c): Security}
  \end{subfigure}
  ~
  \begin{subfigure}[t]{0.48\textwidth}
   \centering
   \includegraphics[height=1.2in]{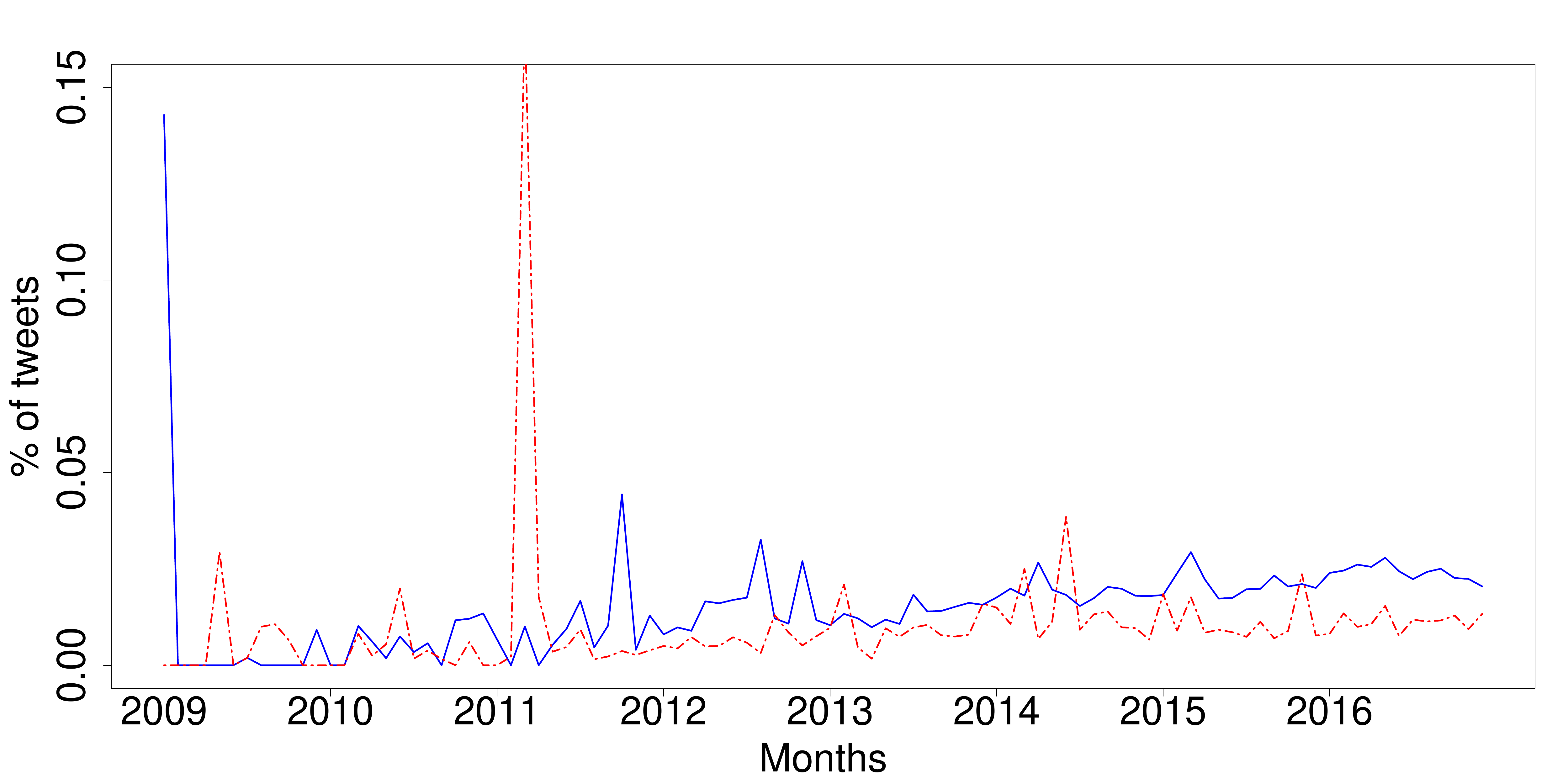}
   \caption{Sentiment for topic (d): Big data \& Tech}
  \end{subfigure}
  \\
  \begin{subfigure}[t]{0.48\textwidth}
   \centering
   \includegraphics[height=1.2in]{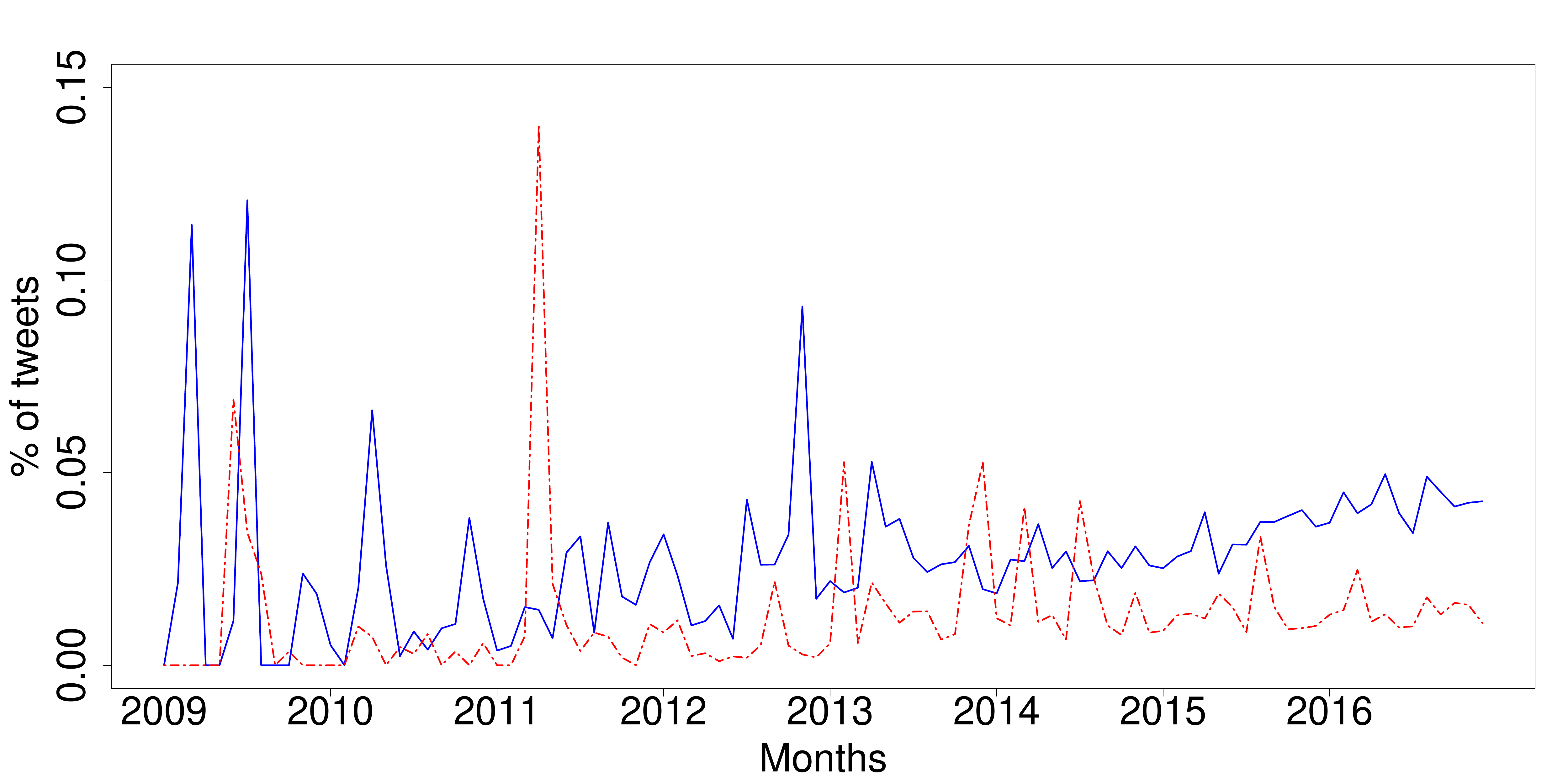}
   \caption{Sentiment for topic (e): Analytics}
  \end{subfigure}
  ~
  \begin{subfigure}[t]{0.48\textwidth}
   \centering
   \includegraphics[height=1.2in]{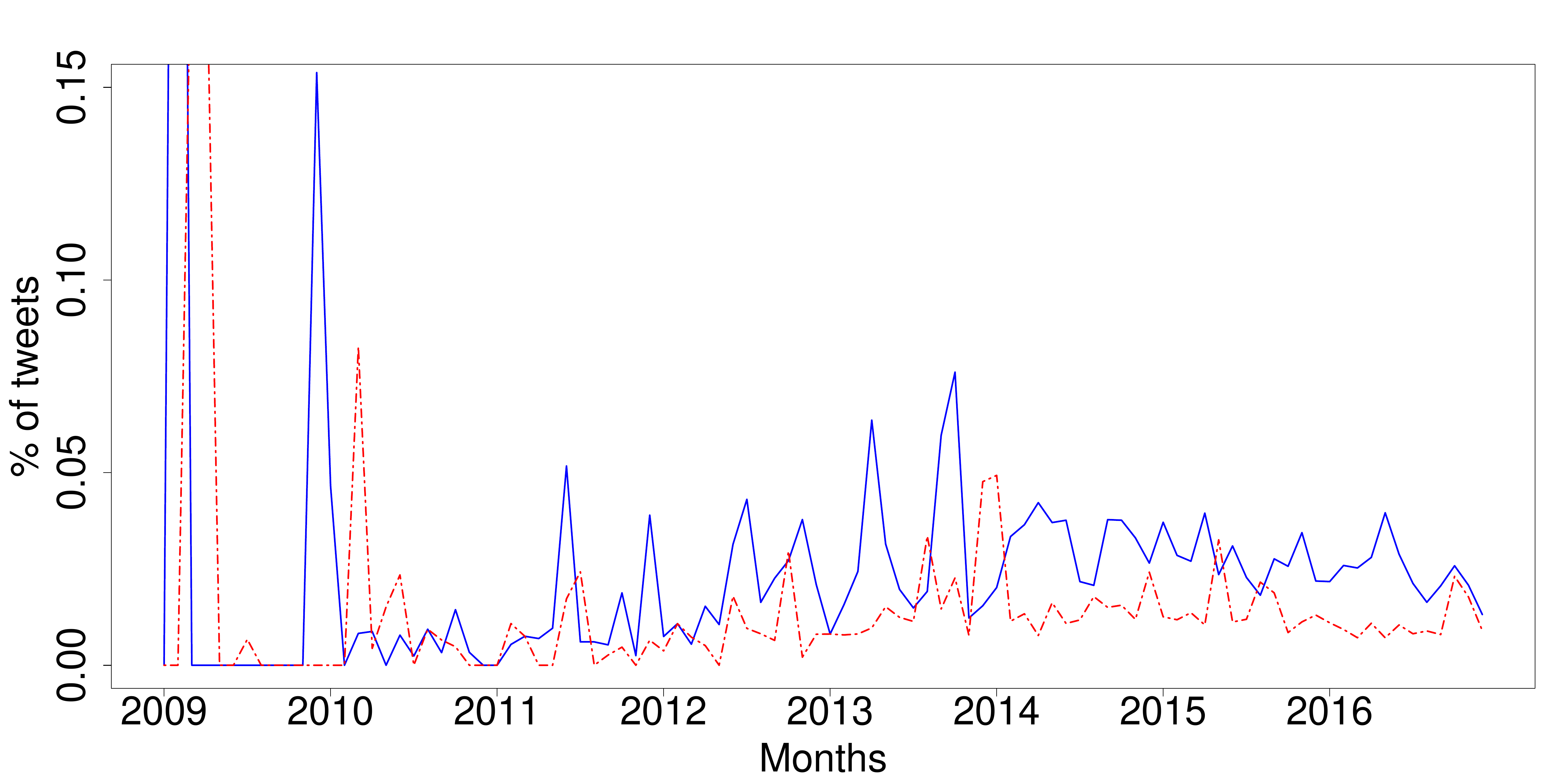}
   \caption{Sentiment for topic (f): Machine learning}
  \end{subfigure}
  \caption{Timeline of sentiments for the 6 topics. Blue, solid line: positive, red, dashed line: negative.}
  \label{fig:sentiment-by-topic}
\end{figure}

Next, we look at the positive and negative sentiment associated with these six topics. Figure \ref{fig:sentiment-by-topic} shows the monthly percentage of positive and negative tweets for each of the topics. In two cases we can observe that positive tweets have increased particularly in recent years, which is the case of ``Analytics'' and ``Machine learning.'' In these two cases, positive tweets clearly exceed negative tweets over the last few years. Two more topics, ``Big data \& Tech'' and ``M2M communication,'' are showing a similar tendency, however, the difference is not as remarkable. `Devices' is a topic that has people divided with similar amounts of positive and negative tweets, and the most negative topic is ``Security,'' which shows an increasing tendency towards more negative tweets in recent years.

It may be surprising, at first thought, to observe that \textit{Analytics}, \textit{Machine learning} and \textit{Big data \& Tech} are viewed as generally positive over the period considered. However, a number of recent surveys have indicated more positive opinions about artificial intelligence and machine learning. The Royal Society report, Public views of Machine Learning \cite{castell2017public}, considered existing work on public attitudes about emerging technologies and found that ``broadly speaking, the public are supportive of science and scientific developments and want to know more about them.'' The report itself considered the current and near-term (5-10 years) applications of machine learning. The research involved 978 face-to-face interviews weighted to ensure the individuals selected for interview were representative of the UK population, and was supplemented with qualitative research. Though the report failed to identify that positive sentiments to machine learning significantly outweighed negative feelings, headlines in the mainstream media included ``Artificial intelligence survey finds UK public broadly optimistic'' \cite{sample2017artificial}.  Academic work has found stronger support for AI and Machine learning, including work by Fast and Horvitz of Stanford University and Microsoft, respectively. Their work found that ``discussion of AI has increased sharply since 2009, and that these discussions have been consistently more optimistic than pessimistic'' \cite{fast2017long}. However, the positivity seen in these algorithmic, software-based aspects of the Internet of Things does not extend to \textit{Devices} or \textit{Security}. The relatively less positive opinion of Devices might be explained by the public being concerned about hardware, and in particular its ability to act, rather than the more ``hidden'' software layer.  It is not surprising that security is an area that is viewed, in aggregate, as negative. Security is one of the key concerns in the Internet of Things, and it is usually only discussed in the media, both mainstream and social, when an incident happens due to a security vulnerability.

It may also be surprising to note that there were generally more tweets that were positive about M2M than those that were negative. This might be explained by considering previous research that explored public opinions of robotics, a significant use case of M2M technology. A recent survey \cite{castell2014public} found that 81\% of respondents favoured the use of robotics in manufacturing.  As such, it is less surprising that there is a slight overall positive attitude to M2M in our results.

If we look at the overall positivity of topics (Table \ref{tab:positivity-by-topic}), we see that the opportunities that the IoT is bringing to business are most positively perceived by people, including \textit{Analytics}, \textit{Machine learning}, \textit{Big data \& Tech} and \textit{M2M communication}. The topics where the negativity exceeds the positivity include especially \textit{Security}, which proves to be the most concerning issue linked to IoT, followed by \textit{Devices} with a lower degree of negativity.

\begin{table}
  \centering
  \begin{tabular}{l l}
   \toprule
   \textbf{Topic} & \textbf{Positivity} \\
   \midrule
   (e) Analytics & +113.02\% \\
   \midrule
   (f) Machine learning & +75.11\% \\
   \midrule
   (d) Big data \& Tech & +53.10\% \\
   \midrule
   (b) M2M communication & +53.07\% \\
   \midrule
   (a) Devices & -5.14\% \\
   \midrule
   (c) Security & -15.09\% \\
   \bottomrule
  \end{tabular}
  \caption{Positivity by topic, computed as the percentage of positive tweets that exceed negative tweets. Topics are sorted by positivity, most positive first.}
  \label{tab:positivity-by-topic}
\end{table}

\subsection*{Countries}

\begin{figure}[tbh]
  \begin{center}
   \includegraphics[width=1.0\textwidth]{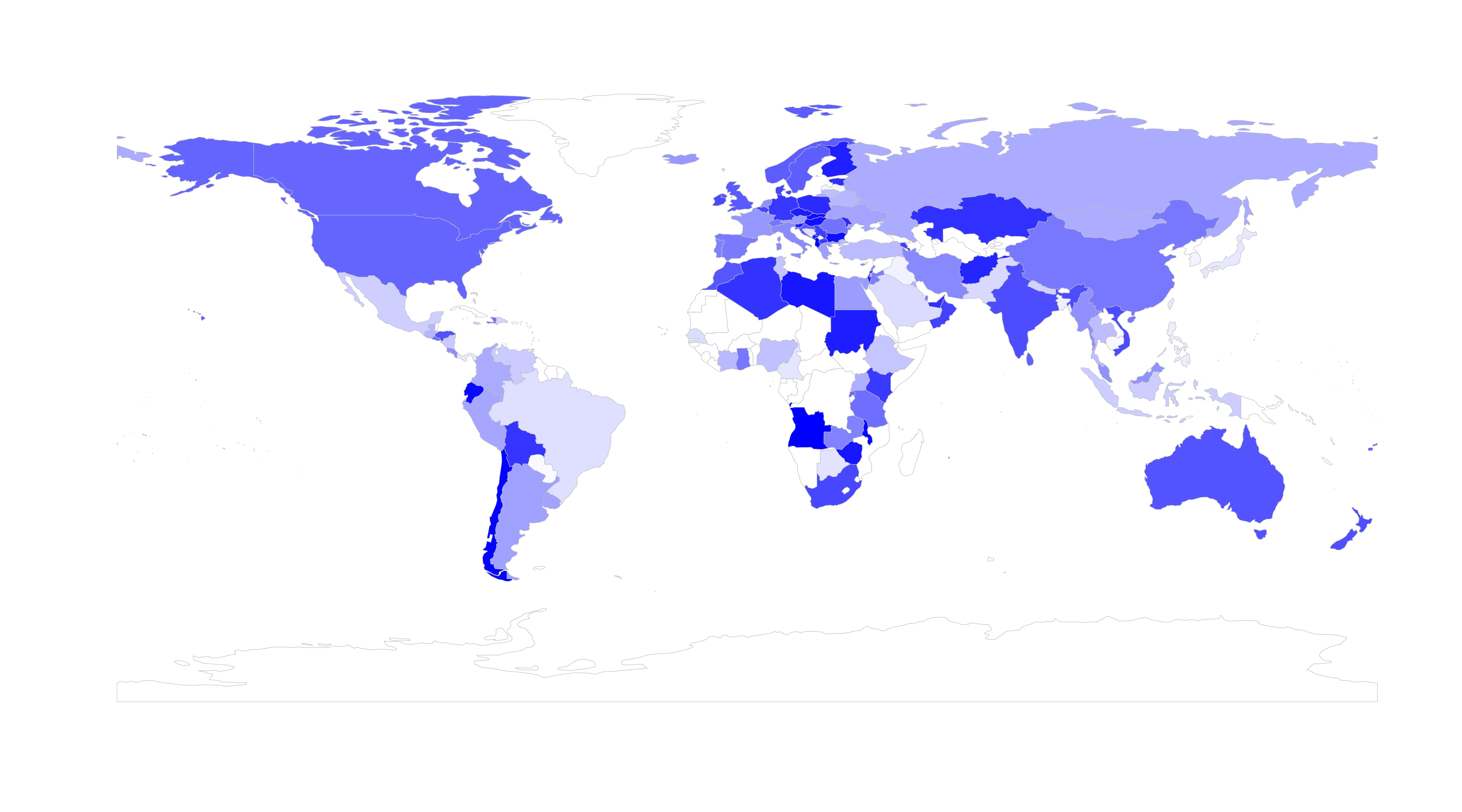}
   \caption{Tweets bearing positive sentiment, by country. Darker blue refers to more positive tweets.}
   \label{fig:sentiment-country-pos}
  \end{center}
\end{figure}

\begin{figure}[tbh]
  \begin{center}
   \includegraphics[width=1.0\textwidth]{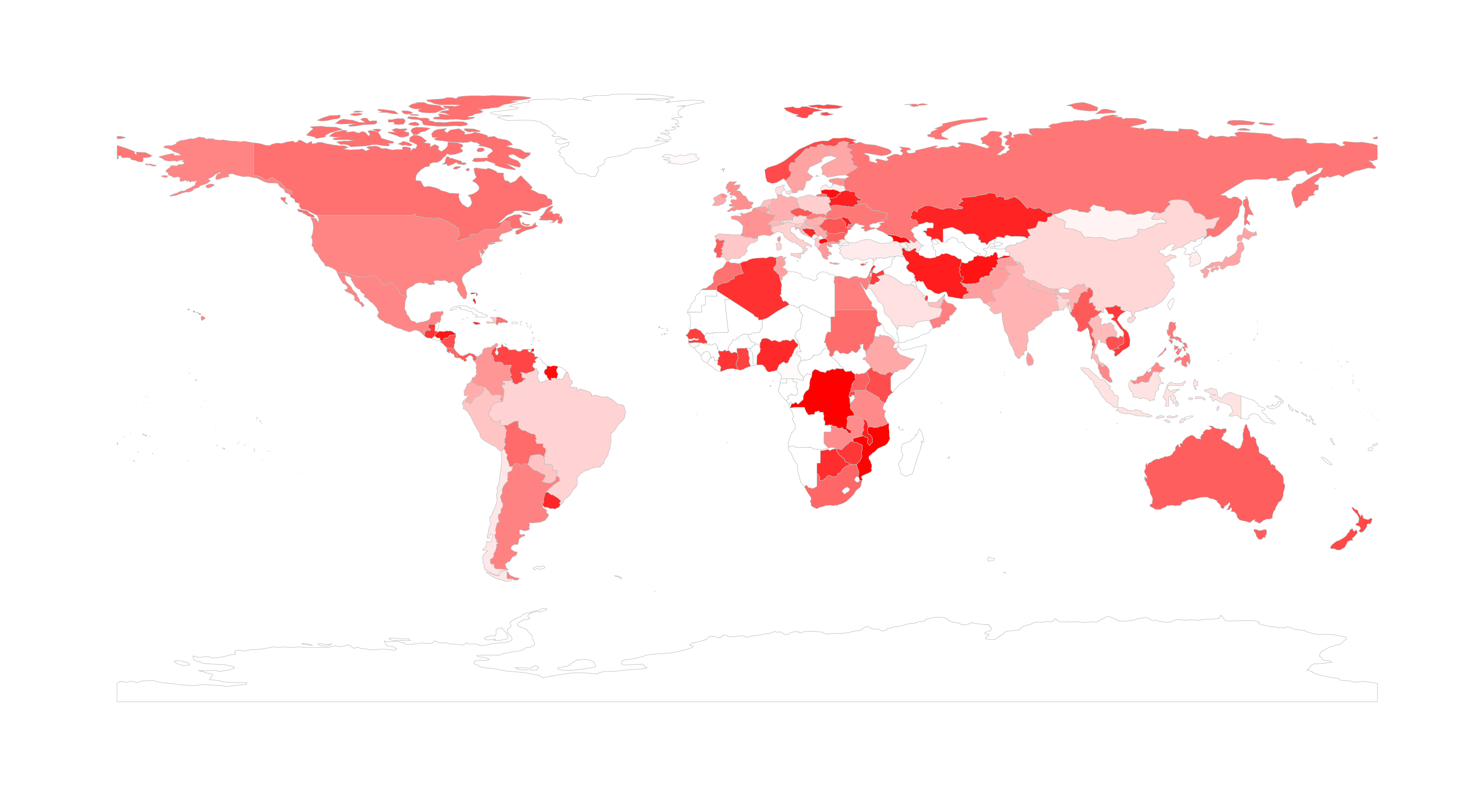}
   \caption{Tweets bearing negative sentiment, by country. Darker red refers to more negative tweets.}
   \label{fig:sentiment-country-neg}
  \end{center}
\end{figure}

By having tweets geo-located by country, we were able to analyse sentiment by country towards the IoT. Figure \ref{fig:sentiment-country-pos} shows a heat map highlighting the extent to which tweets posted (in English) from a specific country are positive, where the countries with the highest percentage of positive tweets are coloured in darker blue. Likewise, Figure \ref{fig:sentiment-country-neg} shows the extent to which tweets posted from a country are positive, with countries with more negative tweets in darker red.

We focus our analysis on the top 10 most active countries in terms of their number of tweets, and look at their positivity, computed as the percentage of positive tweets that exceed negative tweets. The top 10 most active countries are listed in Table \ref{tab:top-countries}, ranked by overall positivity. The table also shows the positivity expressed by these countries towards the two most popular topics, i.e. ``Big data \& Tech'' and ``Security''. All of the top 10 countries show, overall, more positivity than negativity, as shown by the positive values in the `overall' column. This positivity varies substantially, however, from over 92\% for the most positive country, Germany, to a 16.5\% for the least positive country, France. The positivity also varies by topic, with an overall tendency to be more positive towards ``Big data \& Tech'' across different countries, and an overall tendency to be more negative towards ``Security.'' Germany stands out as one of the most positive countries, especially for ``Big data \& Tech.'' The most positive country towards ``Security'' is, however, Spain. While all of the countries are clearly positive towards ``Big data \& Tech,'' six of the top 10 countries are predominantly negative towards ``Security,'' especially Australia, Canada and France, followed by the United Kingdom and the United States, and to a lesser extent by the Netherlands.

\begin{table}
  \centering
  \begin{tabular}{l l l l}
   \toprule
   &  & \textbf{Big data} &  \\
   & \textbf{Overall} & \textbf{\& Tech} & \textbf{Security} \\
   \midrule
   Germany & +92.84\% & +162.28\% & +16.21\% \\
   Indonesia & +72.32\% & +130.85\% & +15.29\% \\
   Spain & +64.01\% & +55.46\% & +30.01\% \\
   Italy & +62.17\% & +77.36\% & +11.29\% \\
   United Kingdom & +46.32\% & +90.14\% & -15.68\% \\
   United States & +42.16\% & +96.08\% & -18.48\% \\
   Netherlands & +38.02\% & +79.75\% & -5.02\% \\
   Canada & +26.21\% & +59.42\% & -28.04\% \\
   Australia & +23.04\% & +62.63\% & -32.39\% \\
   France & +16.52\% & +44.26\% & -26.75\% \\
   \bottomrule
  \end{tabular}
  \caption{Top 10 most active countries, ranked by overall positivity towards the IoT.}
  \label{tab:top-countries}
\end{table}

\begin{figure}[tbh]
  \begin{center}
   \includegraphics[width=1.0\textwidth]{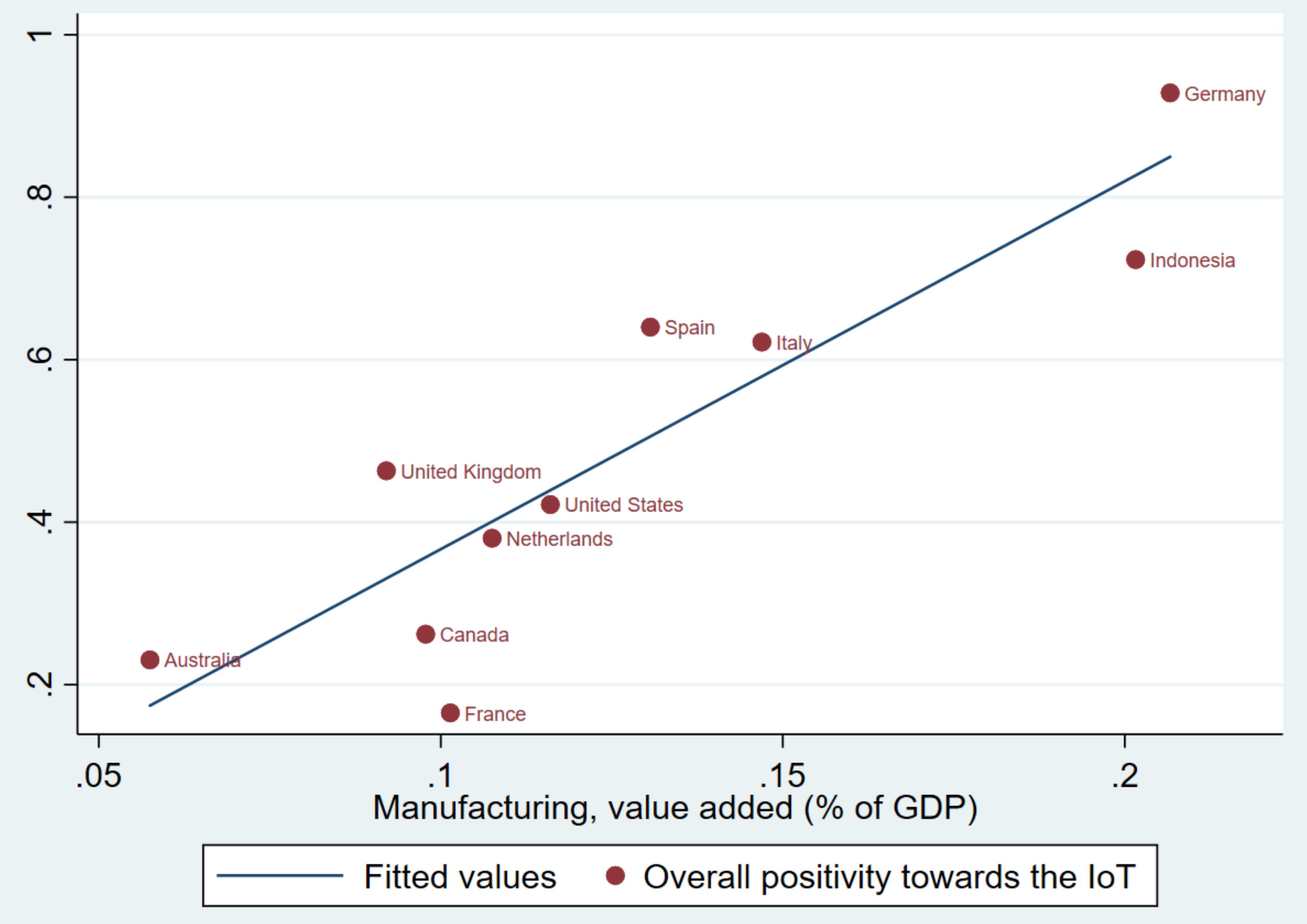}
   \caption{Overall Positivity towards the IoT, by value added from manufacturing.}
   \label{fig:value-added}
  \end{center}
\end{figure}

Understanding why a country such as Indonesia has a similar net positive view of IoT like Germany, whereas its neighbour, France, has a much different view may seem challenging.  However, one explanation that has emerged is that positivity to IoT might reasonably be thought to correlate with a country’s readiness for Industry 4.0. To examine this we have looked at the contribution to a country's GDP provided by manufacturing.  The reason for considering this is that is recognised that some countries ``have embraced the fourth industrial revolution more quickly and fully than many of their counterparts'' because ``manufacturing has consistently been seen as a critical part of those nations' economies'' \cite{schreiber2017comparing}.  We collected statistics of the measurement from the World Bank\footnote{\url{https://data.worldbank.org/}}, and the details of the statistics for the Top 10 most active countries is illustrated in Figure \ref{fig:value-added}.\footnote{We used the latest yearly data available for each of the Top 10 most active countries, which is 2014 for Canada, 2016 for United States and 2017 for all other eight countries. We tested also the results based on latest year which all countries data are available (the year 2014), which provides a similar result.} Accepting that the contribution to a country’s GDP provided by manufacturing is representative of its Industry 4.0 readiness, we can infer from Figure \ref{fig:value-added}, that there is a positive correlation between people’s overall positivity towards the IoT and the country’s readiness to adopting Industry 4.0. This conjecture is confirmed by an additional statistical test, showing the correlation coefficient between the two factors is +0.889. Furthermore, if we assume that the readiness of a country is the causal factor for the overall positivity towards the IoT, a one percent increase in how ready a country to adopting IoT changes is estimated to lead a 5.52 percent significant increase to the overall positivity (p = 0.001).

\subsection*{Gender}

\begin{figure}[tbh]
  \begin{center}
   \includegraphics[width=1.0\textwidth]{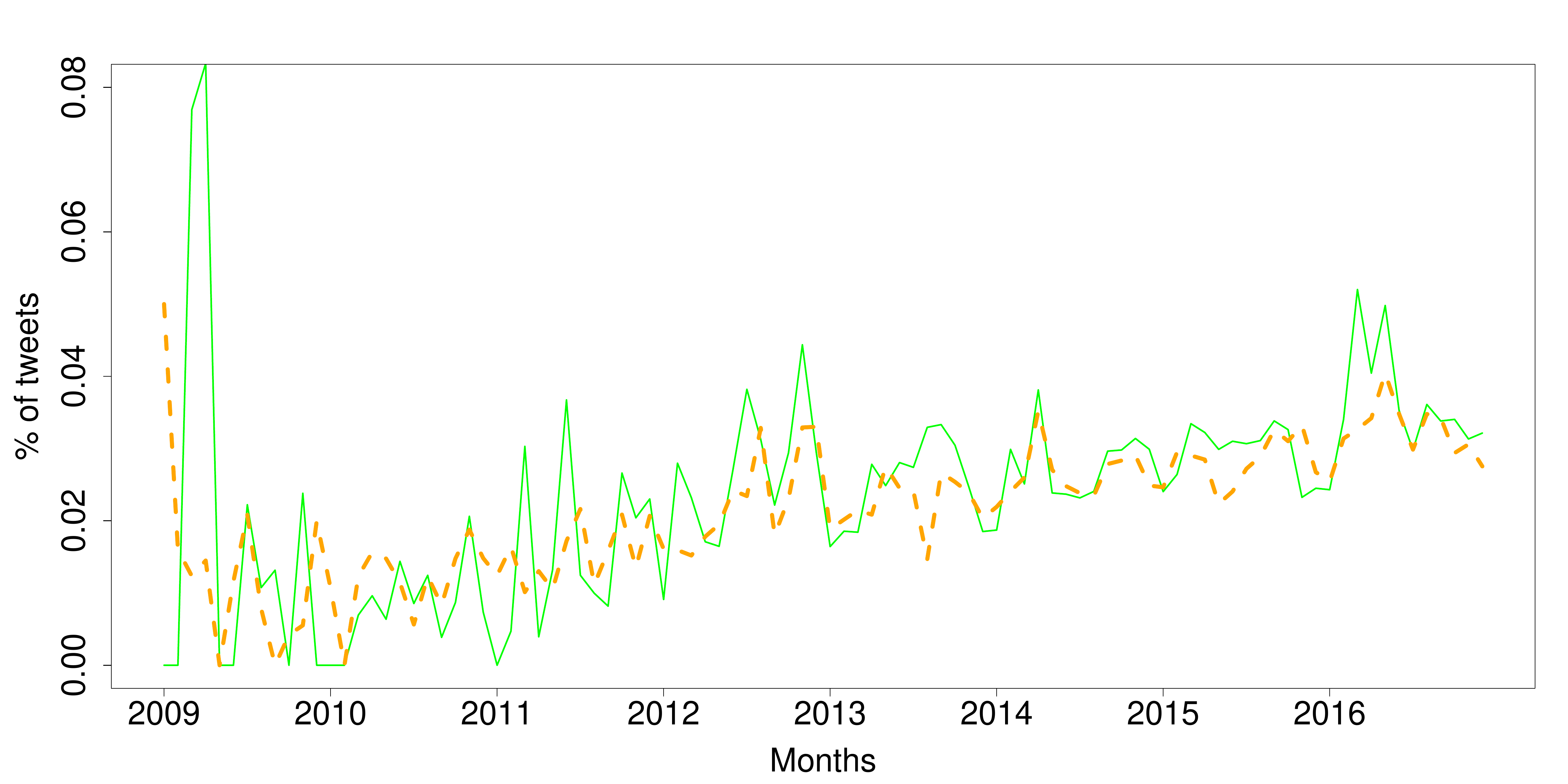}
   \caption{Tweets bearing positive sentiment, broken down by gender (green, solid line: female, orange, dashed line: male).}
   \label{fig:sentiment-gender-pos}
  \end{center}
\end{figure}

\begin{figure}[tbh]
  \begin{center}
   \includegraphics[width=1.0\textwidth]{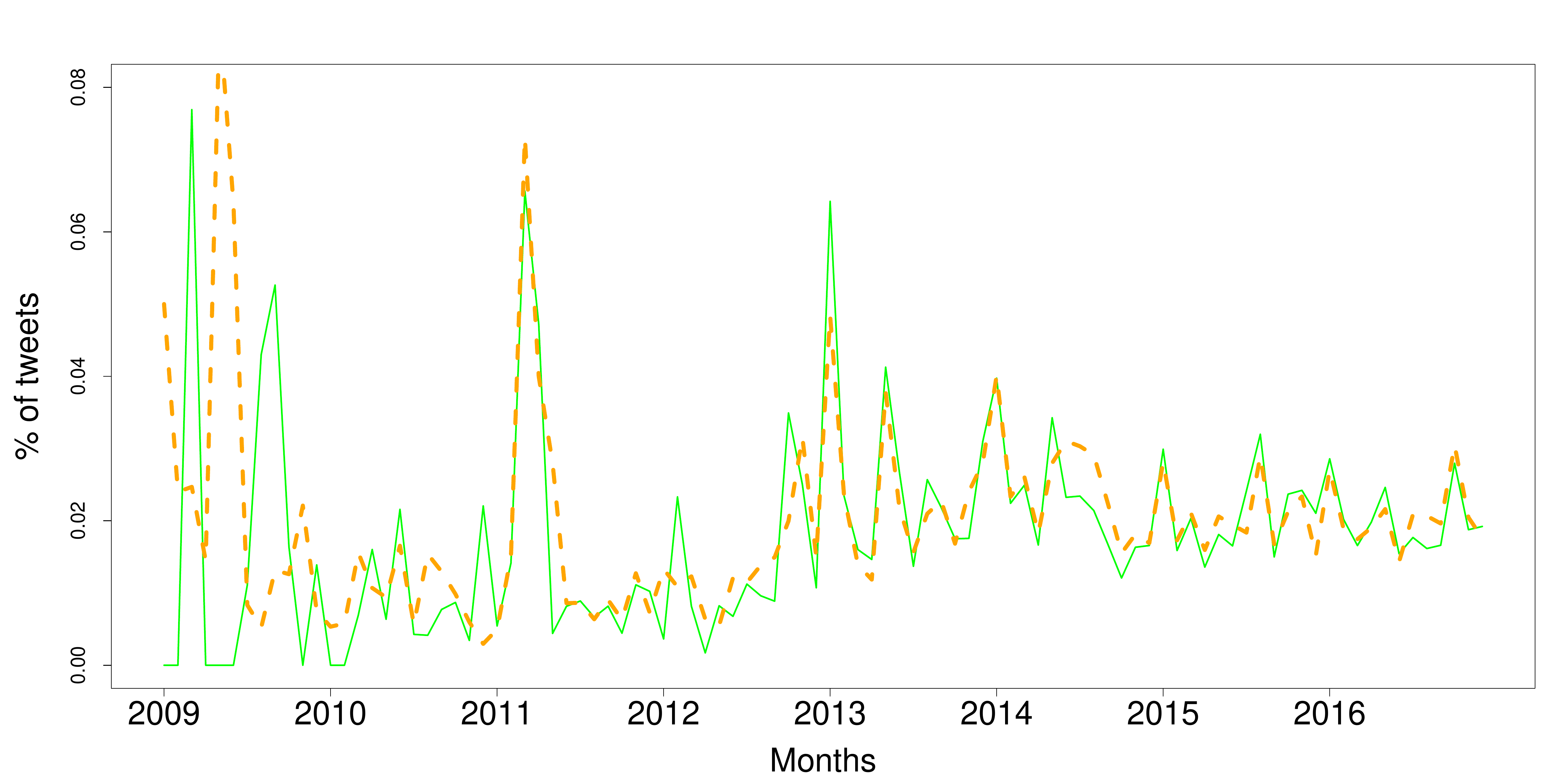}
   \caption{Tweets bearing negative sentiment, broken down by gender (green, solid line: female, orange, dashed line: male).}
   \label{fig:sentiment-gender-neg}
  \end{center}
\end{figure}

Processing of the tweets to identify those posted by accounts that could be identified as being male or female based on their first name led to a subset of 2,602,138 tweets. Given the imbalance in the number of tweets by gender (male: 1,999,934, female: 602,204), we normalised the figures for positive and negative tweets by the number of tweets posted by gender. Figures \ref{fig:sentiment-gender-pos} and \ref{fig:sentiment-gender-neg} show the trends in positivity and negativity for both males and females. Here we show the ratio of positive, neutral and negative tweets posted by males and females. While the charts show an overall similar tendency, an overall comparison of sentiment ratios reveals that, on average, females 9\% more likely to share positive comments about IoT than males, whereas males are 10\% more likely to express negativity, where percentages represent the differences between ratios for each gender. If we analyse the ups and downs in the trends, however, there is some correlation ($\rho = 0.48$ in the positive tweets and $\rho = 0.35$ in the negative tweets). Despite the similarity in the trends for both genders, the overall tendency is for females to be more positive about the IoT. While some studies have shown that males generally are more positive about technology than females, there is previous research on gender attitudes to technology may explain our findings. Almost 20 years ago a study found that, contrary to earlier studies, females ``held more positive attitudes than males regarding the value of computers to make users more productive'' \cite{ray1999men}. Obviously technology has significantly changed in 20 years, and while there has been no significant research into the impact of gender on opinion of the Internet of Things, a recent study that investigated a new model for self-management of psychological stress, based upon the use of an app found that ``females rated the application more positive overall than [...] male participants'' \cite{wiederhold2014marketing}.

\subsection*{News Stories}

News stories can play an important role in shaping public opinion. Our analysis of news stories associated with IoT tweets was based on the external links that users point to in their tweets. As described above, we collect the final URL for all the links in tweets, as many of the links are shortened. In this analysis, we examined the sentiment associated with different links in tweets. First, we looked at the top news stories with the highest number of positive or negative tweets by year, which we show in Table \ref{tab:top-news}. Again, a look at the top news stories suggested that the positive news stories were largely associated with the business opportunities that the IoT provides, whereas negative news were predominantly about security issues, devices being hacked, discussion on the public being scared by the IoT and the risks of the inevitable use of the cloud for the correct functioning of IoT devices.

The data shown in Figure \ref{fig:sentiment-by-topic} show that there are a number of spikes, showing a higher percentage of positive or negative tweets. An example is the publication of the Wired article ``The Internet Of Things Has Arrived -- And So Have Massive Security Issues'' that was released 11 January 2013, during the Consumer Electronics Show, CES 2013 \cite{rose2013internet}. Looking closely at the tweets for that month, a lot of them related to this article highlighting concerns. This was a contributing factor to the spike in negativity observed in January 2013.

\begin{table}
  \centering
  \begin{tabular}{l | p{7cm} | p{7cm}}
   \toprule
   \textbf{Year} & \textbf{Positive} & \textbf{Negative} \\
   \midrule
   2009 & Internet of Things (IoT) - When do you think it will become a business reality? & Are you scared of the Internet of Things? Do RFID chips keep you awake at night, in unholy fear? \\
   \midrule
   2010 & That's it, somebody has gone and coined ``Web 3.0 - The Internet of Things'' - ...let the insanity begin. :/ & This internet of things can be dangerous - hacker disables 100 cars remotely \\
   \midrule
   2011 & Inspiring The Internet Of Things: A Comic Book: The Internet of Things is one of our favorite trends at RWW. & How the Internet of Things is Changing the Way We Work \\
   \midrule
   2012 & Enjoying the internet of things? Thank your smartphone. & A french startup is disrupting the biz - Does the internet of things need its own internet? \\
   \midrule
   2013 & WiFi Bunnies and Why I Love the Internet of Things & Why the internet of things has to be not too smart and not too dumb, but just right \\
   \midrule
   2014 & The internet of things is great for chipmakers and a challenge for Intel & Why the Internet of Things narrative has to change. \#Internet\_of\_things. \\
   \midrule
   2015 & IoT is on the cusp of something great, do you have a strategy in place? & \#InternetofThings to cause major \#security headaches \\
   \midrule
   2016 & \#IoT-enabled devices with connectivity to \#SAP is a great example of innovation in \#IT & Why a Marriage Between the Cloud and Internet of Things Is Inevitable \\
   \bottomrule
  \end{tabular}
  \caption{Top news headlines bearing positive or negative sentiment by year.}
  \label{tab:top-news}
\end{table}

\section*{Discussion and Conclusions}

We have performed a large-scale, longitudinal analysis of the public perception of the different opportunities and challenges presented by the Internet of Things (IoT), which we have achieved by mining data from the social media platform Twitter. We have used the topic modelling algorithm LDA to identify the six main topics discussed by the public around the IoT. We analysed the polarity of tweets by using a state-of-the-art target-specific sentiment analysis algorithm, and by further exploring other dimensions such as country of origin and gender.

Among the six topics we identified in the dataset, the two most popular topics include ``Big data \& Tech'' and ``Security.'' This reveals that despite the business interest that the IoT presents for big data analytics, the challenges posed by the limited security of today's IoT devices are a major concern for the general public. We further confirm this with the predominantly negative sentiment that is associated with posts discussing security issues. This is again further confirmed when we look at the top news stories shared each year, with negative news being predominantly about security issues associated with IoT devices. Our study raises awareness on the importance of keeping IoT devices secure, reminding manufacturers that it is a concern that is being continually discussed.

A finer-grained analysis shows, however, variations across countries. While some countries like Germany, Indonesia and Spain tend to be generally positive about the IoT, others such as the United Kingdom, United States, Australia and France are not as positive, with a significant negative tendency towards security issues. We also observed some differences across gender groups, women being 9\% more likely positive about IoT and men being 10\% more likely negative.

The main finding that may impact on the large scale adoption of IoT is the public concern associated with security and IoT, suggesting that further effort is needed on the part of IoT device and service providers to convince the public that the IoT is -- or will be -- secure. \cite{li2017securing,yager2017new}.

\section*{Future work}

As is the case with Twitter users in general, our dataset is not a representative sample of the general population, hence the results cannot be treated as being a true reflection of public opinion. To address this issue, we are planning to use additional information extraction techniques to help determine key socio-demographic variables, such as age and education \cite{rao2010classifying}.

We also intend to use the findings of this work to inform further qualitative and quantitative research in a mixed-paradigm research study \cite{sale2002revisiting} on perceptions of the IoT, following established principles \cite{venkatesh2013bridging}. Clearly since Twitter posts are restricted in length, it is difficult to ascertain in depth information on the thoughts being presented.  Using this work we intend to explore further, through focus groups, why there is a difference in perception between Analytics and Big Data and Devices and Security. Analysis of these workshops will allow us to formulate a nationally representative survey to determine national perceptions on IoT.  This work would build upon our analysis and seek to find answers to open questions identified in the literature, such as \cite{asplund2016attitudes}.

Furthermore, the tools we have used are independent of the topic and thus applicable to other datasets of English tweets. In the future, we aim to collect other datasets to analyse public opinion on a range of issues of public interest, including political issues such as the United Kingdom's decision to leave the European Union (Brexit) and major societal issues such as abortion.

% Do NOT remove this, even if you are not including acknowledgments.
\section*{Acknowledgements}
This work is part of the PETRAS project, which is funded by a grant from the UK Engineering and Physical Sciences Research Council (EP/N02334X/1).

\nolinenumbers

\bibliographystyle{vancouver}
\bibliography{iot}

\end{document}